\documentclass[preprint,12pt]{elsarticle}

\usepackage{graphicx}

\usepackage{amssymb}

\usepackage{lineno}

\begin{document}

\begin{frontmatter}



\title{Performance of the ATLAS Muon Drift-Tube Chambers at High Background Rates and in Magnetic Fields}


\author{J.~Dubbert}
\author{S.~Horvat}
\author{F.~Legger}
\author{O.~Kortner\corref{cor1}}
 \ead{Oliver.Kortner@mpp.mpg.de}
\author{H.~Kroha}
\author{R.~Richter}
\author{Ch.~Valderanis}
 \address{Max-Planck-Institut f\"ur Physik, F\"ohringer Ring 6,
    80805 Munich, Germany}

\author{F.~Rauscher}
\author{A.~Staude}
 \address{Ludwig-Maximilians-Universit\"at M\"unchen, Schellingstr. 4,
    80799 Munich, Germany}

\cortext[cor1]{Corresponding author}

\begin{abstract}
The ATLAS muon spectrometer uses drift-tube chambers for precision tracking. The performance of these chambers in the presence of magnetic field and high $\gamma$~radiation fluxes is studied in this article using test-beam data recorded in the Gamma Irradiation Facility at CERN. The measurements are compared to detailed predictions provided by the Garfield drift-chamber simulation programme.
\end{abstract}


\end{frontmatter}


\section{Introduction}
\label{introduction}

The ATLAS detector will be used to study proton proton collisions at large centre-of-mass energies of up to 14~TeV at the Large Hadron Collider (LHC). It is equipped with a muon spectrometer for efficient muon identification and precise transverse momentum measurement with a resolution of 4\% up to $p_T\approx100$~GeV/c and better than 10\% up to $p_T\leq1$~TeV/c. The ATLAS muon spectrometer consists of a system of superconducting air-core toroid coils generating a magnetic field delivering field integrals of about 2.5~Tm in the barrel part and up to 7~Tm in the end caps. The muon trajectories in the muon spectrometer are measured with three layers of so-called monitored drift-tube (MDT) chambers whose shapes and relative positions are monitored by a system of optical sensors with micrometer accuracy. The MDT chambers are built of two quadruple layers of cylindrical drift tubes in the innermost layer and of two triple layers of drift tubes in the remaining two layers. The tubes have an outer diameter of 30~mm and a 400~$\mu$m thick tube walls made of aluminium. They are filled with a gas mixture of argon and carbon dioxide in the mixing ratio of 93:7 at an absolute pressure of 3~bar. The anode wire of 50~$\mu$m diameter is set to a voltage of 3080~V with respect to the tube walls leading to a gas gain of 20,000 and an average spatial single-tube resolution better than 100~$\mu$m. This corresponds to a spatial resolution of the chambers of better than 40~$\mu$m as required for 10\% momentum resolution at 1~TeV/c \cite{ATLAS_det_paper}.

The magnetic field in the muon spectrometer is non-uniform, also within the muon chambers. The magnetic field $B$ influences the electron drift inside the tubes: the maximum drift time $t_{max}\approx700$~ns increases by approximately 70~$\frac{\mathrm{ns}}{\mathrm{T^2}}B^2$. $B$ varies by up to $\pm0.4$~T along the tubes of chambers mounted near the magnet coils which translates into a variation of $t_{max}$ with an amplitude of 45~ns \cite{MPI_2}. The dependence of the space drift-time relationship must be taken into account in track reconstruction. Test-beam measurements presented in this article were needed to model this dependence with the required accuracy of 1~ns \cite{ELBA}.

The operating conditions of the MDT chambers at the LHC design luminosity of 10$^{34}$~cm$^{-2}$s$^{-1}$ are characterized by unprecedentedly high neutron and much higher $\gamma$ background fluxes. The chambers are expected to show background counting rates of up to 500~s$^{-1}$cm$^{-2}$ including all uncertainties in the prediction of the radiation background \cite{Shupe_1,Shupe_2}. The choice of the Ar:CO$_2$ gas mixture and the use of silicon free components in the gas system prevents the MDT chambers from aging \cite{Aging_1, Aging_2}. The high radiation background is known to degrade the spatial resolution of the tubes and the muon detection efficiency. The performance of the drift tubes at high $\gamma$ irradiation background was studied with single tubes and preliminary read-out electronics for the first time in the late 1990s \cite{Riegler_1}. Measurements with a large standard MDT chamber and the final read-out electronics could only be performed in 2003 \cite{MPI_1}; these measurements are consistent with the earlier measurements using single tubes. They were followed up by measurements of the drift-tube behaviour in magnetic fields in 2004 \cite{MPI_2}.

The present article summarizes the studies published in \cite{MPI_2}, \cite{ELBA}, and \cite{MPI_1} and extends them by detailed comparisons of the measurements with predictions of the Garfield drift chamber simulation programme \cite{Garfield}.

\section{\label{setup}Test-beam set-ups}
\begin{figure}[hbt]
\begin{center}
    \includegraphics[width=\linewidth]{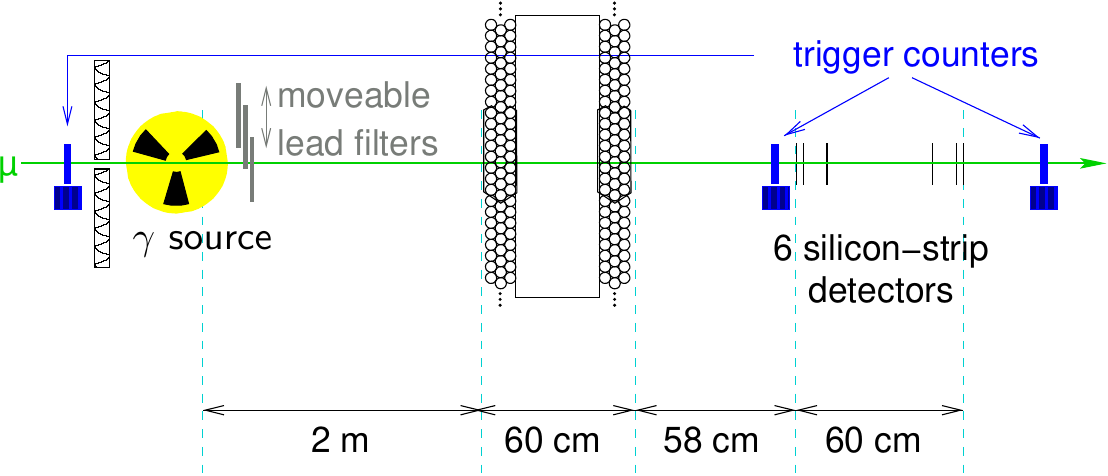}
    \caption{\label{Fig1}Schematic drawing of the experimental set-up in 2003. A full-size MDT chmaber was operated in the GIF.}
\end{center}
\end{figure}

\begin{figure}[hbt]
\begin{center}
    \includegraphics[width=\linewidth]{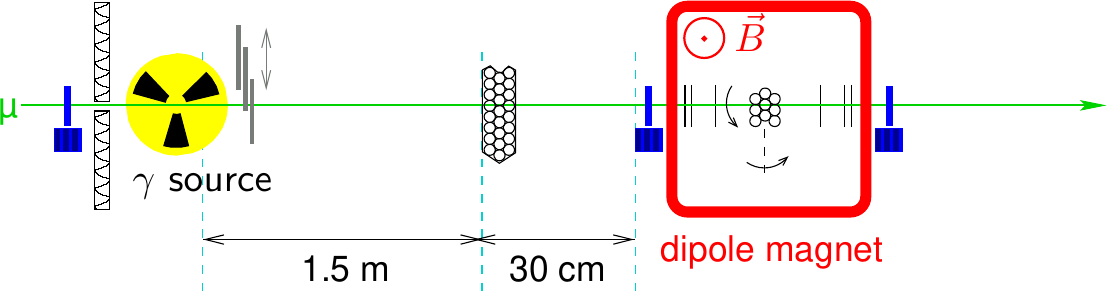}
    \caption{\label{Fig2}Schematic drawing of the experimental set-up in 2004. A small bundle of short drift tubes is mounted on a rotation table in the middle of the beam hodoscope inside a large dipole magnet. A bundle of 3.8~m long drift tubes is installed in front of the source outside the magnet for high-rate studies.}
\end{center}
\end{figure}

The measurements discussed in this article were made at the Gamma Irradiation Facility (GIF) at CERN \cite{GIF} which provided a 90 GeV muon beam and a 664~GBq $^{137}$Cs source emitting 66~keV $\gamma$ rays used to simulate the ATLAS radiation background. The flux of $\gamma$ rays to which the drift tubes were exposed during the measurements could be adjusted with movable lead filters in front of the opening of the source. Two experimental set-ups were installed in the GIF, one in 2003, the other in 2004 -- see Figures~\ref{Fig1} and \ref{Fig2}. Common to both installations are the trigger system and the beam hodoscope.

The beam hodoscope consists of two times three silicon strip detectors of 5$\times$5~cm$^2$ active area 60~cm apart. Each of the four inner strip detectors measures the lateral positions of the incident muons with 10~$\mu$m accuracy. The two outermost detector planes measure the vertical muon positions with the same accuracy. In 2004 one of these vertical detectors had to be turned off due to very high leakage currents.

The beam hodoscope is enclosed by two fast scintillation counters with the same active area as the silicon detectors. A coincidence of these two counters with a 10$\times$10~cm$^{2}$ scintillator counter mounted upstream in the beam and protected from the radiation of the $\gamma$ source provides the trigger and the start time for the drift-time measurements in the tubes. The jitter of the trigger time of less than 0.5~ns is negligible with respect to the time resolution of the drift tubes of about 3~ns.

In 2003 one of the largest ATLAS MDT chambers was operated in the GIF with final read-out electronics. This chamber contains 432 tubes of 3.8~m length in two triple layers separated by 317~mm. It was installed at 2~m distance from the radioactive source. In the final MDT chamber read-out scheme each drift tube is connected to a chain of an amplifier, shaper, and discriminator \cite{Arai} capacitively coupled to the sense wire. A time-to-digital converter measures the arrival times of the signal with respect to the trigger time for each tube in a group of 24. Bipolar shaping is used with at peaking time of 15~ns. The discriminator threshold is adjustable and set to a value corresponding to 5 times the thermal noise created by the 383~$\Omega$ termination resistor on the high-voltage side of the tubes. The pulse heights of the signals are measured by an analog-to-digital converter in each of the 24 channels. Bipolar shaping was chosen in order to avoid baseline shifts of the analog signals caused by multiple hits in the presence of high radiation background in the ATLAS muon spectrometer. A consequence of bipolar shaping are multiple threshold crossings of the discriminator threshold. The secondary threshold crossings can be masked by an artificial dead time of 790~ns after the detected hit \cite{MPI_2}. This is the baseline for the operation in ATLAS at the LHC. The dead time is adjustable between 200~ns and 790~ns. While we used the baseline value of 790~ns in 2003, we also used the minimum dead time of 200~ns in 2004.

In 2004 a bundle of 24 3.8~m long drift tubes was installed at 1.5~m distance from the source to study the drift tube performance with the minimum dead time of the electronics. The other important goal of the 2004 test-beam programme was the precision measurement of the influence of the magnetic field on the space drift-time relationship of the tubes. For this measurement, a chamber with 3$\times$3 15~cm long ATLAS drift tubes in the middle of the beam hodoscope was operated in a 0-0.9~T dipole field measured with a Hall probe on the chamber. The orientation of the magnetic field vector $\vec{B}$ with respect to the direction of the anode wires of the tubes could be changed by rotating the chamber inside the magnet.

\section{Drift-tube properties in the absence of radiation background}
In this section, the space drift-time relationship, the spatial resolution, the pulse-height spectra, and the drift-tube efficiency will be discussed for the operation of the MDT chambers in the absence of $\gamma$ radiation background. We first present our studies for the operation without magnetic field and then the analysis of the measurements with magnetic field.

\subsection{\label{sec31}Operation with no magnetic field}
Throughout the analysis muon trajectories are reconstructed with the data of the beam hodoscope independently of the muon chambers. The high spatial resolution of the silicon strip detector planes of the hodoscope leads to a precise knowledge of the muon position inside the drift tubes with a resolution much better than the spatial resolution of the drift tubes.

\begin{figure}[hbt]
\begin{center}
    \includegraphics[width=0.5\linewidth]{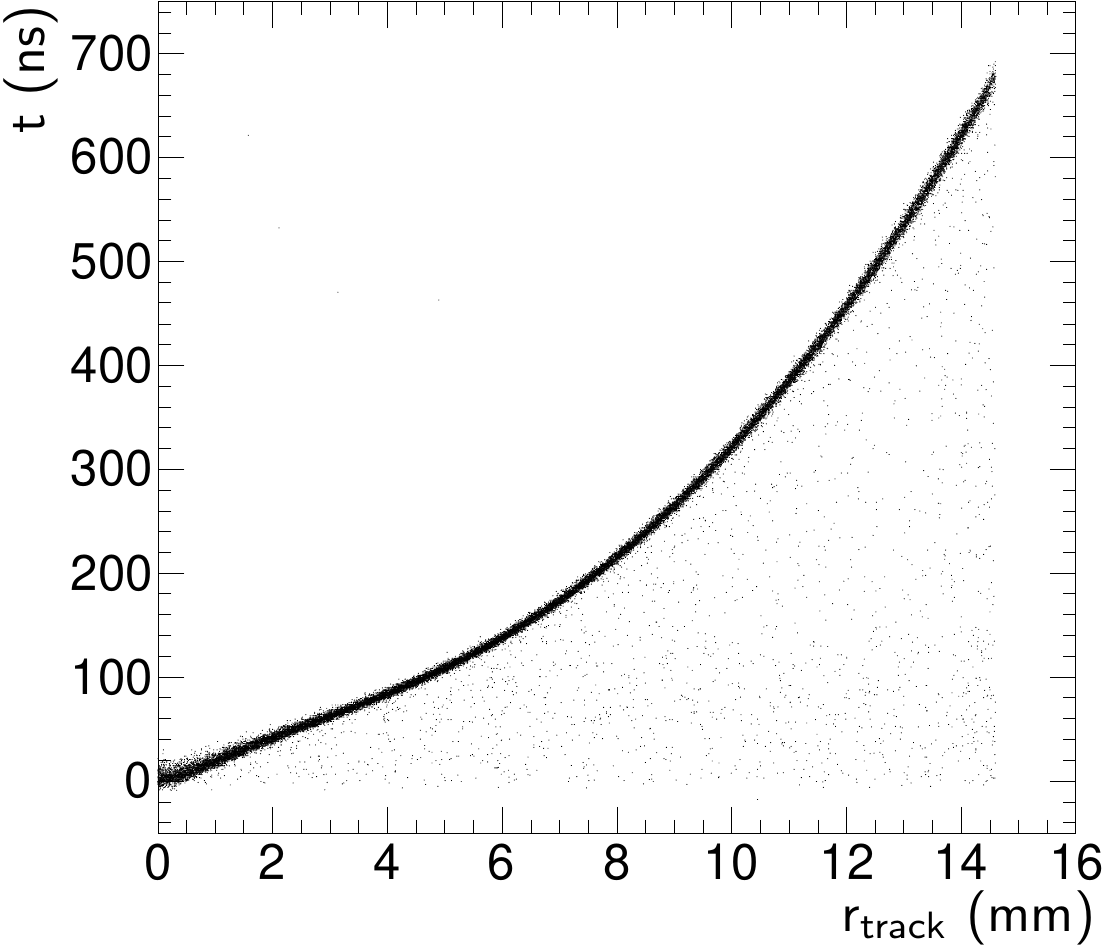}
    \caption{\label{Fig3}Drift times measured by a tube traversed by muons as a function of the distance of the muons from the tube's anode wire as reconstructed by the beam hodoscope. The MDT read-out electronics is operated with the maximum dead time of 790~ns.}
\end{center}
\end{figure}

Figure~\ref{Fig3} shows the drift time measured by a tube traversed by the muons as a function of the distance of the reconstructed hodoscope trajectory from the anode wire of the tube. Most of the data points populate a dark band whose contour reflects the space drift-time relationship. The space drift-time relationship is approximately linear for radii less than 5~mm with an average drift velocity of 0.05~mm\,ns$^{-1}$. The drift velocity drops continously with increasing radius above 5~mm and has a mean value of about 0.017~mm\,ns$^{-1}$ there. The width of the dark band is a measure for the drift time resolution which is of the order of a couple of nanoseconds. 7\% of the data points lie below the dark band. These entries are caused by $\delta$ electrons which are knocked out of the tube walls by the incident muons leading to hit times before the muon hits.

\begin{figure}[hbt]
\begin{center}
    \includegraphics[width=0.5\linewidth]
                                {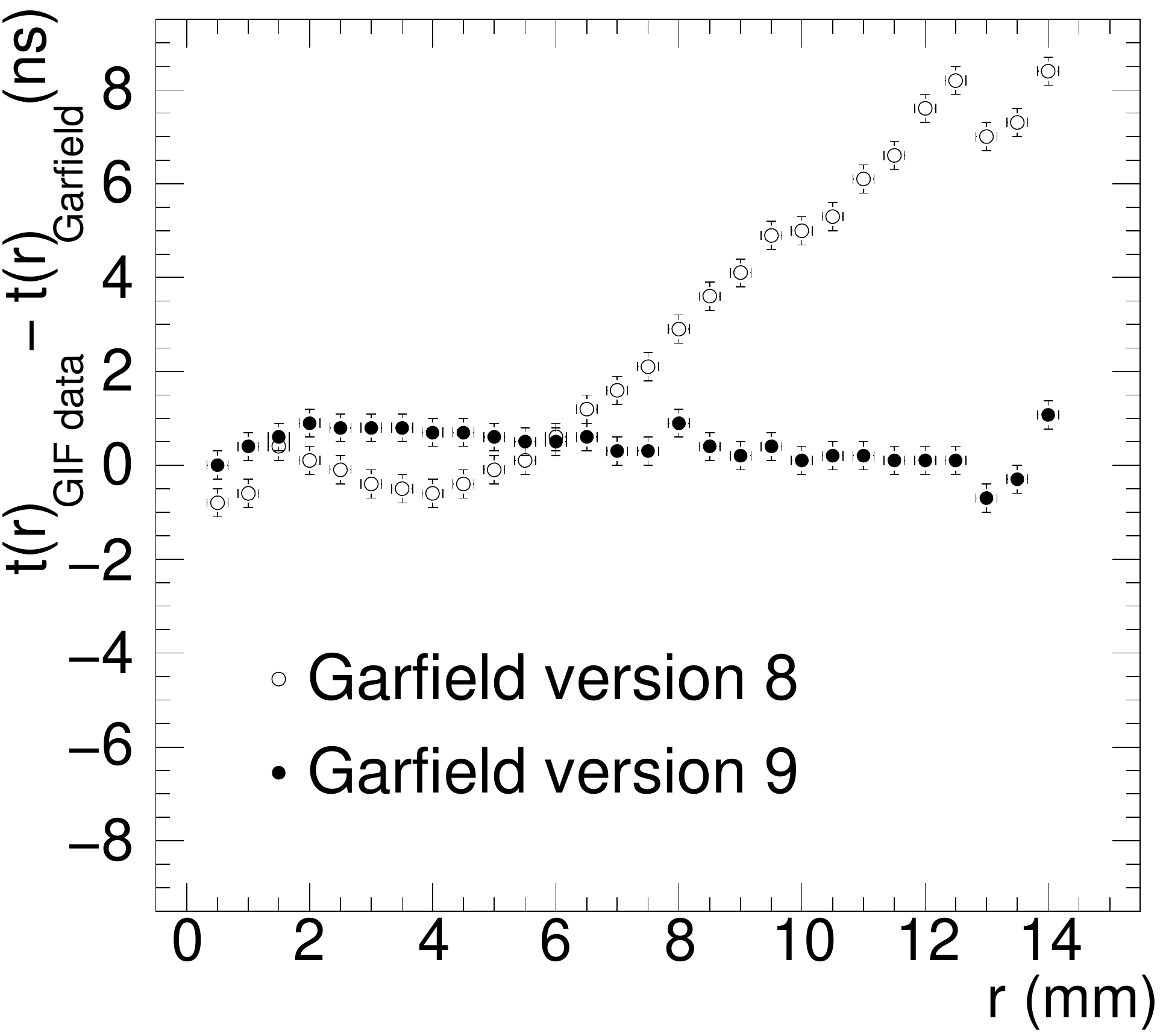}
    \caption{\label{Fig4}Comparison of the measured space drift-time relationship $t(r)_{GIF~data}$ with the predictions $t(r)_{Garfield}$ of the two versions of the Garfield drift-chamber simulation programme.}
\end{center}
\end{figure}

We begin the detailed analysis of the data with a comparison of the measured space drift-time relationship to the prediction of the drift-chamber simulation programme "Garfield". We use Garfield versions 8 and 9. The main and most important difference between the two versions are the cross sections for the collisions of the drifting electrons with the carbon dioxide molecules in the chamber gas. In the cross sections provided to Garfield-8 by the Magboltz-2 programme \cite{Magboltz2} rotational excitations of CO$_2$ molecules by impinging electroncs are neglected. They are taken into account by Magboltz-7 \cite{Magboltz7} used in Garfield-9. Figure~\ref{Fig4} shows the deviation of the measured space drift-time relationship $t(r)_{GIF~data}$ from the space drift-time relationships predicted by the two Garfield versions as a function of the distance of the muon trajectory from the anode wire. The prediction by Garfield-9 agreed with the measured space drift-time relationship on the level on 1~ns. Garfield-8, however, predicts a too small drift velocity for the electroncs leading to deviations $t(r)_{GIF~data}-t(r)_{Garfield}$ of up to 8~ns at large radii. This is caused by the underestimation of the electron-CO$_2$ cross section due to the neglection of rotational excitations of the CO$_2$ molecules whose energy levels are of the same order as the kinetic energy of the drifting electrons. One can make the Garfield-8 prediction match the measured space drift-time relationship by increasing the CO$_2$ content from its correct value of 7\% to 7.08\%. Yet also after this tuning Garfield-8 fails to correctly describe the magnetic-field dependence of the space drift-time relationship as we shall see in the next section.

Unless stated otherwise, Garfield-9 is used in the analysis of our test-beam data from now on.

\begin{figure}[hbt]
\begin{center}
    \includegraphics[width=\linewidth]{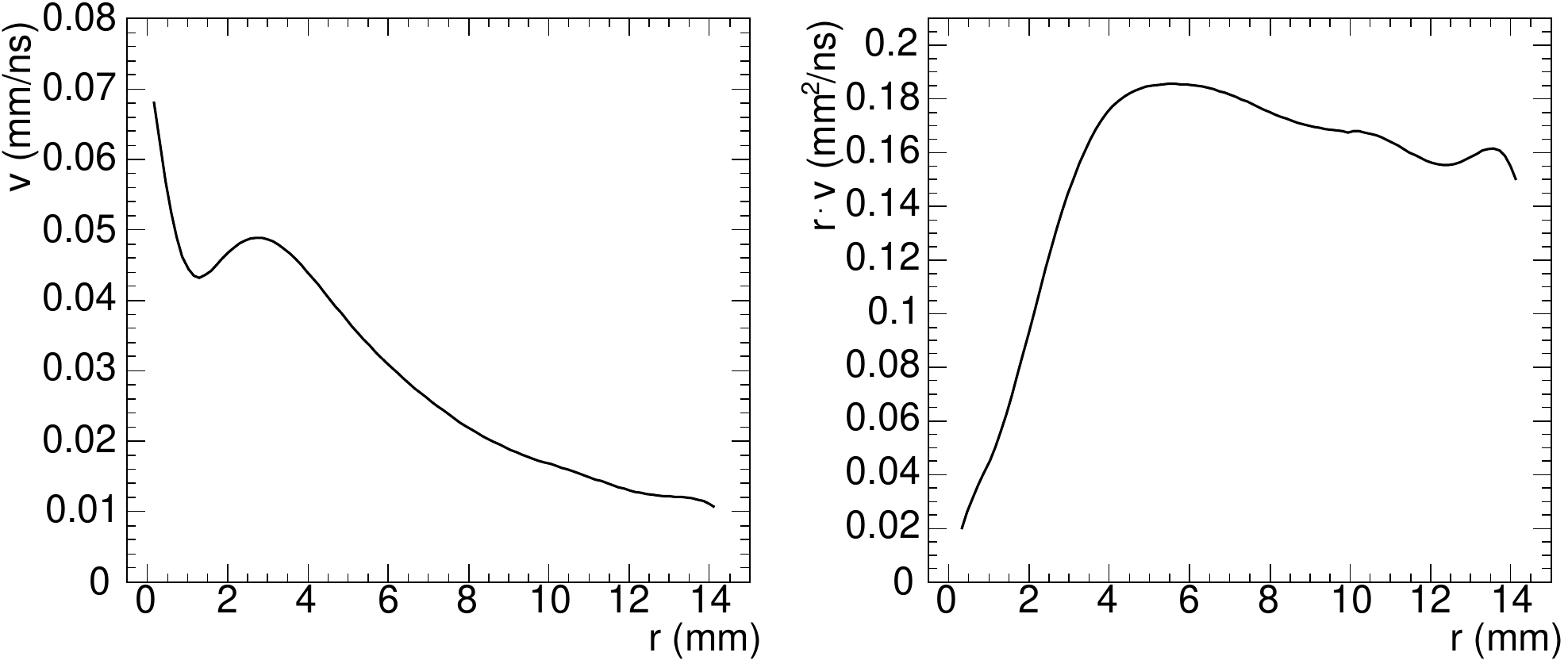}
    \caption{\label{Fig5}Left: Electron drift velocity $v$ as a function of the drift radius $r$. Right: $r\cdot v$ to illustrate the radial dependence of the drift velocity.}
\end{center}
\end{figure}

Before investigating the spatial resolution of the drift tubes we shall briefly discuss the drift velocity of the electrons. Figure~\ref{Fig5} shows the electron drift velocity as a function of $r$ as obtained by differantiating the nineth order polynomial fitter to the contour of the dark band in Figure~\ref{Fig3}. As mentioned above, the drift velocity is about 0.05~mm\,ns$^{-1}$ for $r\lesssim5$~mm. It has a minimum at $r=1$~mm which is caused by the so-called Ramsauer minimum in the cross section for the elastic scattering of electrons on argon atoms. The drift velocity decreases with increasing radius for $r\gtrsim5$~mm. On the right hand side of Figure~\ref{Fig5} we plot $r\cdot v(r)$ as a function of $r$ to illustrated the radial dependence of the drift velocity $v(r)$: $r\cdot v(r)$ is rising linearly for $r\in[0,~4~\mathrm{mm}]$ reflecting that $v(r)$ can be approximated by a constant. At larger radii $r\cdot v(r)$ is roughly constant which is equivalent to the statement that $v\propto \frac{1}{r}$ for $r\gtrsim5$~mm to certain approximation. We shall use the approximations of the radial dependence of the drift velocity to explain the shape of the tube resolution in the next paragraph.

\begin{figure}[hbt]
\begin{center}
    \includegraphics[width=0.5\linewidth]{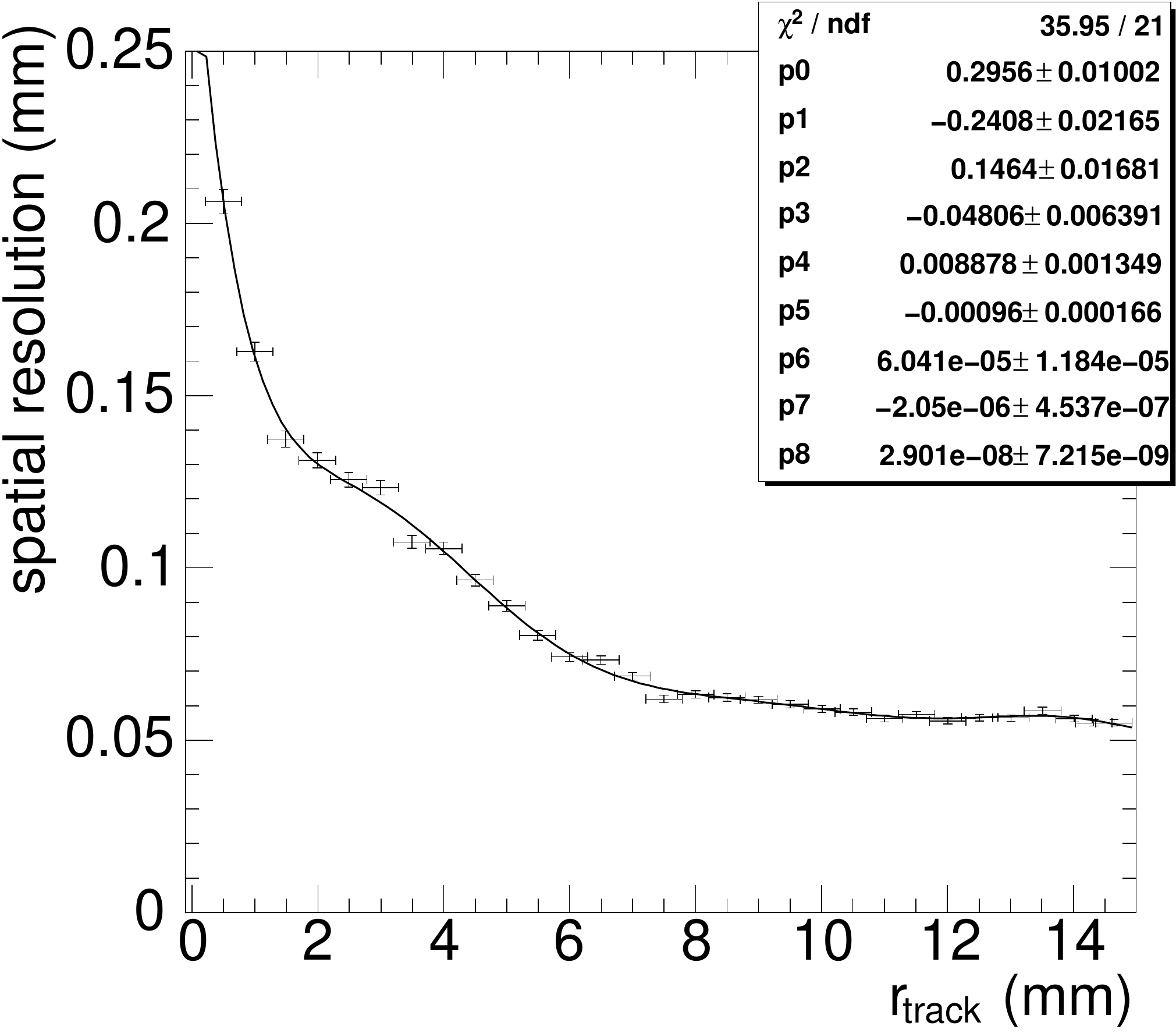}
    \caption{\label{Fig6}Spatial resolution $\sigma_r(r_{track})$ as a function
    of the track impact radius $r_{track}$. A polynomial of 8th order
    $\sigma_r(r_{track})=\sum\limits_{k=0}^{8}p_k r_{track}^k$ is fitted to the
    data points resulting in the given parameters.}
\end{center}
\end{figure}

We define the spatial resolution of the drift tube at a given radius $r_{track}$
as the standard deviation of the normal distribution fitted to the
$r(t)-r_{track}$ distribution where $r(t)$ denotes the drift radius measured by
the tube and $r_{track}$ is given by the beam hodoscope. To prevent
$\delta$~electron hits from spoiling the fit a cut on $|r(t)-r_{track}|<1$~mm is
applied before the fit. The measured spatial resolution of the drift tube is
presented in Figure~\ref{Fig6}.
It is about 200~$\mu$m for $r_{track}<1$~mm and improved with increasing track
impact radius down to 55~$\mu$m for $r_{track}\gtrsim7$~mm. A better
understanding of the shape of the spatial resolution $\sigma_r(r_{track})$ can be
obtained by comparing it to the Garfield predictions. The results of this
comparison become more intuitive when one looks at the drift-time resolution
$\sigma_t(r_{track})$ which is related to the spatial resolution by the equation
$\sigma_t(r_{track})=\frac{1}{v(r_{track})}\sigma_r(r_{track})$, $v(r_{track})$
being the drift velocity of the electrons in the Ar:CO$_2$ gas mixture.

\begin{figure}[hbt]
\begin{center}
    \includegraphics[width=0.5\linewidth]{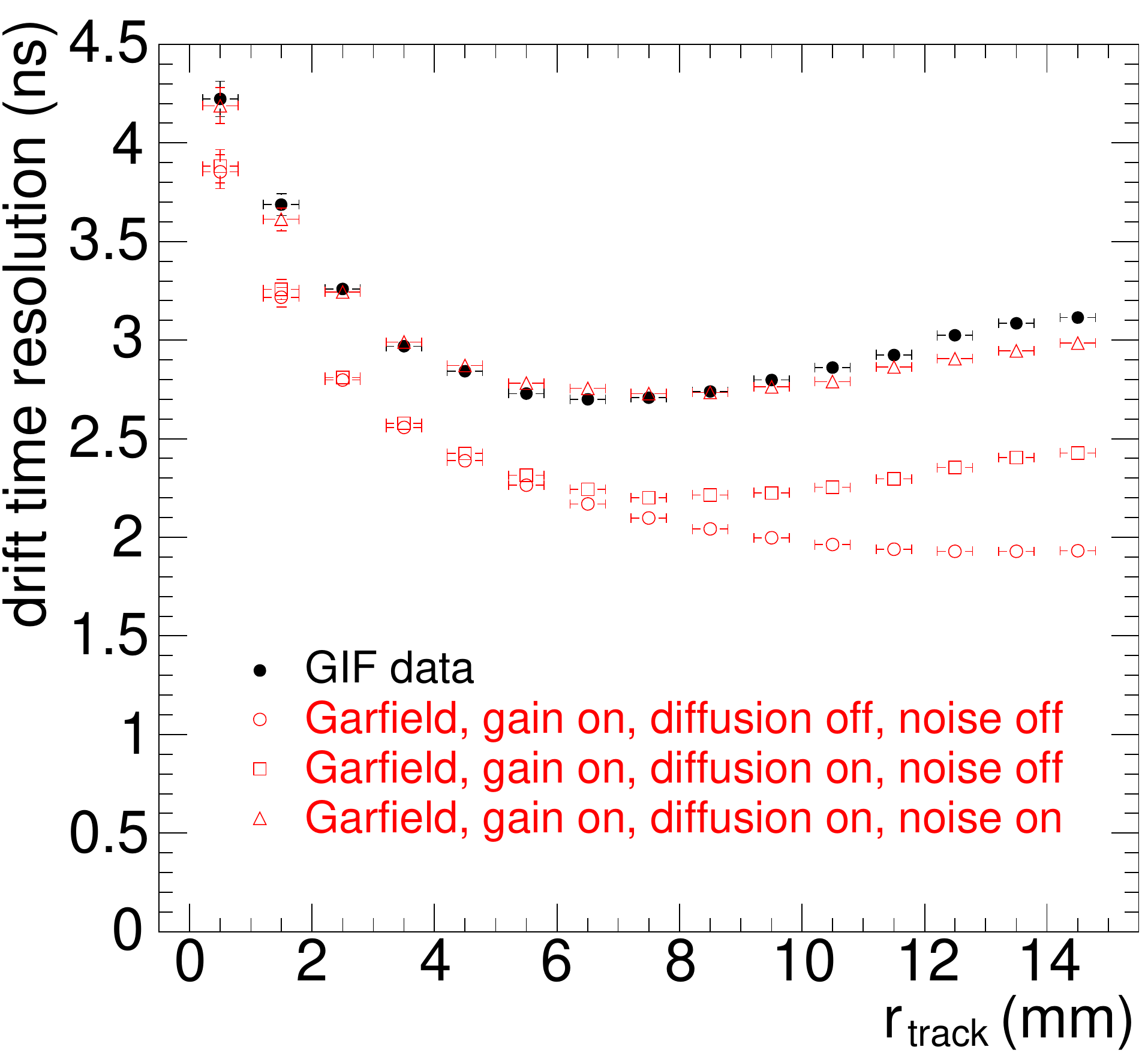}
    \caption{\label{Fig7}Comparison of the measured drift-time resolution with
    Garfield predictions. A discriminator threshold corresponding to 11 times the
    signal of a single primary electron is used in the Garfield simulation. The
    noise level in the simulation is set to $\frac{1}{5}$ of the threshold.}
\end{center}
\end{figure}

Figure~\ref{Fig7} contains three time-resolution curves calculated with Garfield. The open circles show the resolution which is obtained when diffusion in the electron drift and noise are turned off in the simulation programme, but a gas gain of 20,000 is applied. The time resolution improves significantly with increasing distance of the muon track from the anode wire for $r_{track}\lesssim6$~mm and saturates to a constant value for $r_{track}\gtrsim6$~mm. This behaviour can be explained by a simple geometrical model in combination with the radial dependence of the drift velocity of the
electrons. The discriminator threshold is set to a value corresponding to the signal of  $n(=11)$ primary electrons. So the drift-time fluctuation is caused by
the fluctuation of the arrival times of the $n^{th}$ primary electron at the anode wire. Let $s$ be the mean muon track length between $n$ primary ionizations.
Then the $n^{th}$ primary electron travels the distance
\begin{eqnarray}
    r':=\sqrt{r_{track}^2+\frac{s^2}{4}}.
    \nonumber
\end{eqnarray}
The actual distance between $n$ primary ionizations fluctuates around the mean
value $s$ by $\delta s$. Hence the distance the $n^{th}$ primary electrons drifts
fluctuates by
\begin{eqnarray}
    \delta r' = \frac{1}{4}\frac{s}{r'}\delta s.
    \nonumber
\end{eqnarray}
from one muon track to the other. The fluctuation of the arrival times of the
$n^{th}$ primary electrons is given by
\begin{eqnarray}
    \delta t := v(r')\delta r'\
        \approx \frac{1}{4}s\delta s \cdot \frac{v(r_{track})}{r_{track}}
    \nonumber
\end{eqnarray}
where $r'$ has been approximated by $r$. The drift velocity $v(r_{track})$ is constant for $r_{track}\lesssim5$~mm to good approximation, so $\delta t$ falls off like $\frac{1}{r_{track}}$. $v(r_{track}$ is proportional to $\frac{1}{r_{track}}$ for $r_{track}\gtrsim5$~mm making $\delta t$ a constant. This is the behaviour predicted by Garfield when diffusion is turned off. Diffusion causes an increasing fluctuation of the arrival time of the $n^{th}$ primary electron with increasing drift distance. The noise on the signals which is independent of $r_{track}$ shifts the resolution curve upwards onto the measured resolution curve. Only a very small descrepancy of 0.1~ns between the final Garfield prediction from the measured resolution is observed for $r_{track}>12$~mm.

The read-out electronics of the tubes is equipped with an analog-to-digital converter measuring the charge of each amplified and shaped signal within 15~ns after the discriminator threshold has been crossed. The charge measured by the ADC within the integration gate $\Delta t_{gate}$(=15~ns) is proportional to the number of primary ionization electrons arriving at the wire within the time interval $[t(r_{track}), t(r_{track})+\Delta t_{gate}]$. The difference of the arrival times of the electrons freed at $
r_{track}$ and $r'=\sqrt{r_{track}^2+\frac{s^2}{4}}$ -- $\frac{s}{2}$ being the distance of the two electrons on the trajectory of the ionizing muon -- is given by
\begin{eqnarray}
    \Delta t_{arrival}:=\frac{1}{v}(r'-r_{track})
        \approx \frac{1}{v}\frac{s^2}{8r_{track}^4}.
    \nonumber
\end{eqnarray}
Setting $\Delta t_{arrival}$ tp $\Delta t_{gate}$ leads to the maximum separation $s_{max}$ of two primary electrons which can contribute to the charge measurement of the ADC:
\begin{eqnarray}
    s_{max} \approx 2\sqrt{2}\sqrt{\Delta t_{gate}}\sqrt{r_{track}v(r_{track})}.
    \nonumber
\end{eqnarray}
As the charge $q$ measured by the ADC is proportional to the number of primary ionization electrons, $q$ is proportional to $s_{max}$, hence
\begin{eqnarray}
    q \propto \sqrt{r_{track} \cdot v(r_{track})}.
    \nonumber
\end{eqnarray}
\begin{figure}[hbt]
\begin{center}
    \includegraphics[width=0.5\linewidth]{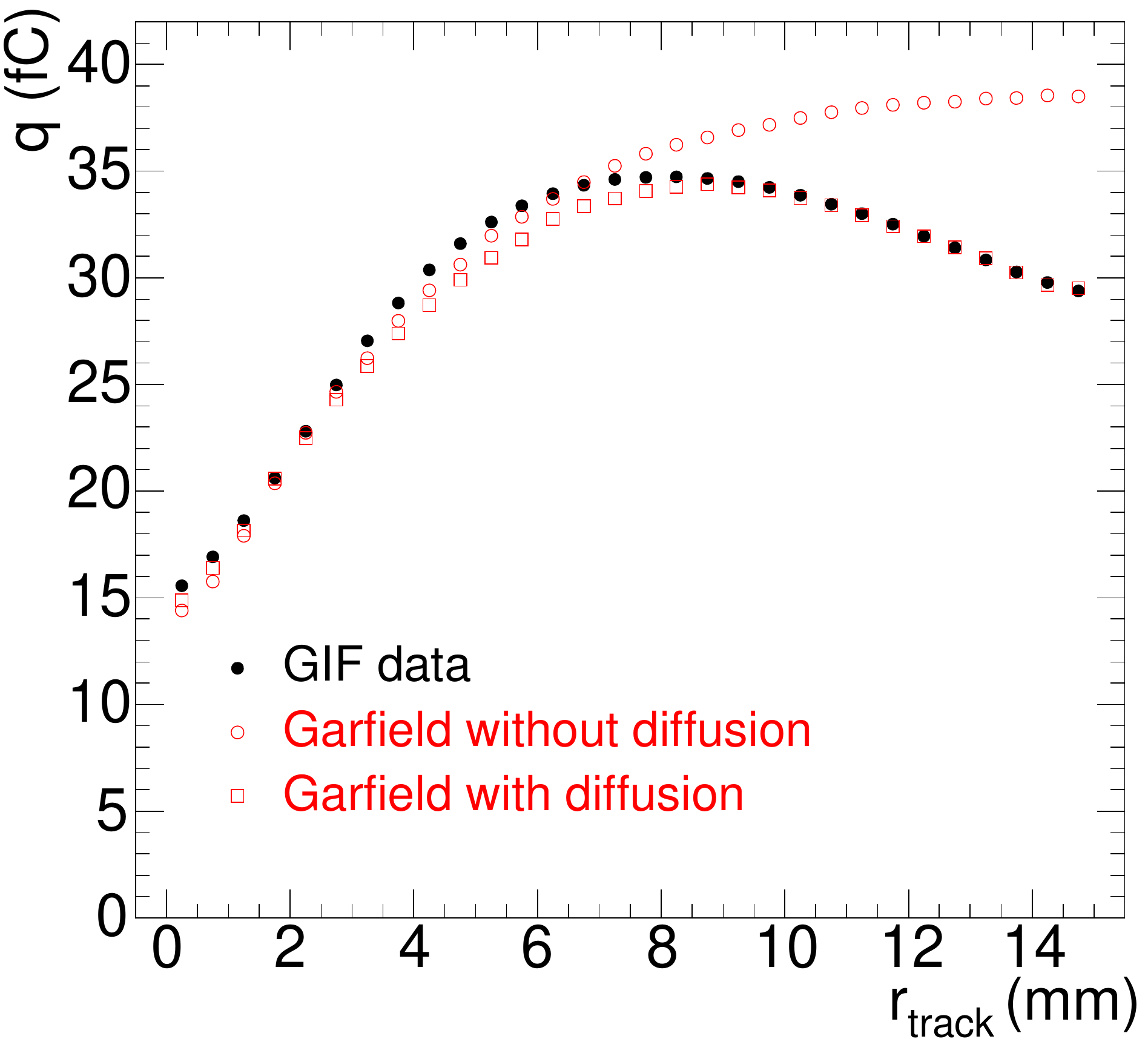}
    \caption{\label{Fig8}Measured and predicted pulse-height distributions as a function of the track impact radius. The plot shows the most probable charge measurement at a given radius.}
\end{center}
\end{figure}
As $v\approx const$ for $r_{track}\lesssim5$~mm, $q$ is expected to increase with increasing radius for $r_{track}\lesssim5$~mm. $r_{track}v(r_{track})\approx const$ for $r\gtrsim5$~mm, so $q$ is expected to saturate at radii $\gtrsim5$~mm. The Garfield prediction of the most probable charge measurement at a given radius presented in Figure~\ref{Fig8} shows such a behaviour when diffusion is turned off. Diffusion enlarges the spread of the arrival times of the primary electrons for large drift distances. As a consequence less primary electrons arrive at the anode wire within the ADC gate and the measured charge decreases with increasing radius for $r_{track}\gtrsim8$~mm. The initial rise of the measured charge for small radii and the drop at large radii are present in the GIF data and well predicted by the Garfield simulation programme.

\begin{figure}[hbt]
\begin{center}
    \includegraphics[width=0.5\linewidth]{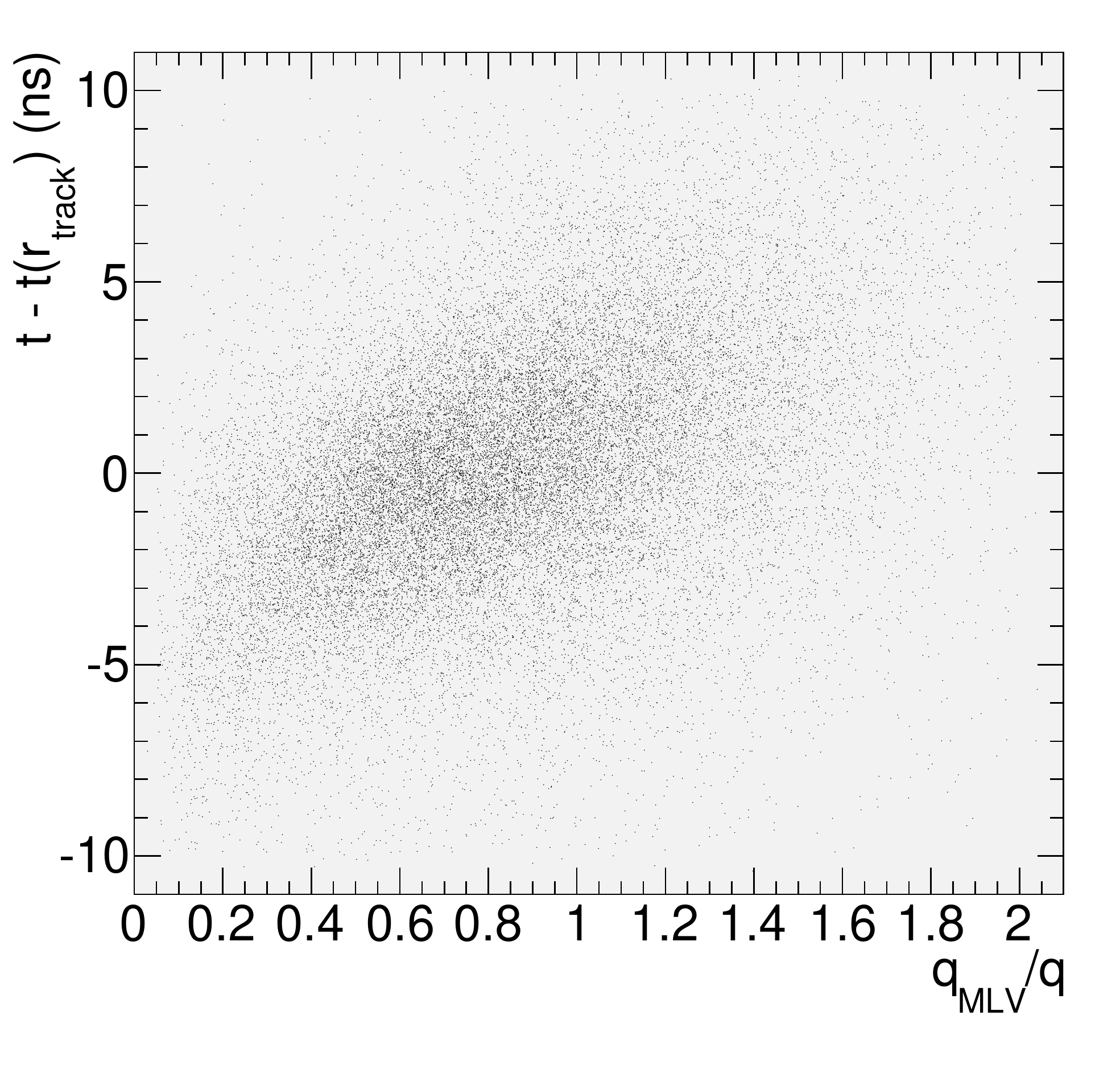}
    \caption{\label{Fig9}Dependence of the deviation of the measured drift time $t$ from the expected drift time $t(r_{track})$ on the ratio of the most likely pulse height $q_{ML}$ and the actual pulse height $q$.}
\end{center}
\end{figure}
\begin{figure}[hbt]
\begin{center}
    \includegraphics[width=0.5\linewidth]{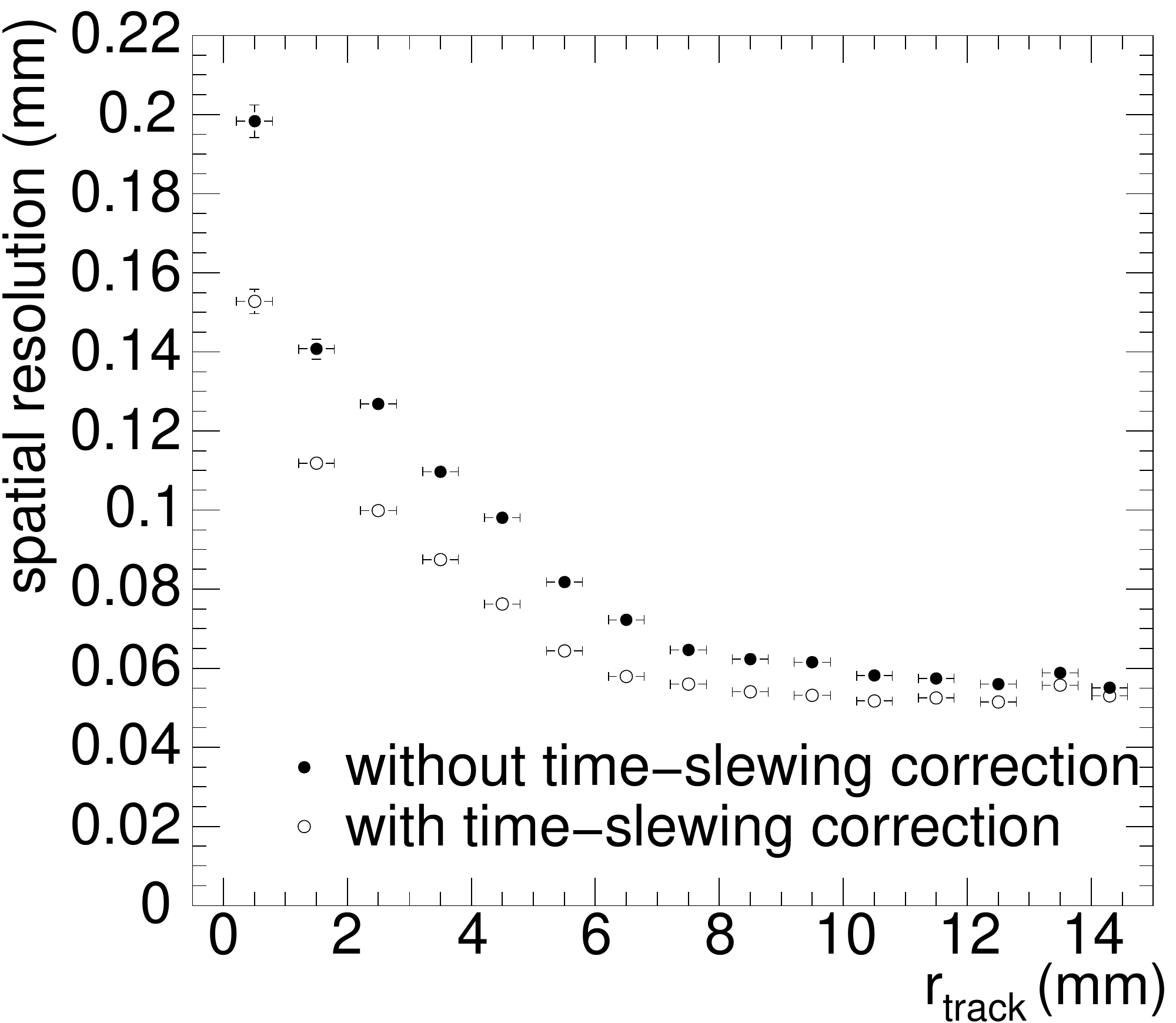}
    \caption{\label{Fig10}Radial dependence of the spatial resolution of a drift tube with and without time-slewing correction.}
\end{center}
\end{figure}

The integration gate of the ADC is of the same order as the peaking time of the shaper. The charge measured by the ADC is therefore proportional to the pulse height of the signal. The knowledge of the height of each signal pulse makes it possible to correct for the walk of the discriminator threshold crossing time with the pulse height. Let $q_{ML}$ denote the most likely height of the signals produced by a muon at a given radius and $q$ be the height of a single pulse. If one approximates the rising edge of the signal by a straight line
\begin{eqnarray}
    q(t)=\frac{q}{\tau} \cdot t
    \nonumber
\end{eqnarray}
where $\tau$ is the peaking time, the walk of the threshold crossing $t_{thr}(q)$ with respect to the crossing time $t_{thr}(q_{ML})$ of a pulse with the most likely height is given by
\begin{eqnarray}
    t_{thr}(q)-t_{thr}(q_{ML})=q_{threshold}\tau
        \left(
        \frac{1}{q}-\frac{1}{q_{ML}}
        \right) =
        q_{threshold}\tau\frac{1}{q_{ML}}
        \left(
        \frac{q_{ML}}{q}=1
        \right),
    \nonumber
\end{eqnarray}
i.e. proportional to $\frac{q_{ML}}{q}$. The dependence of the deviations of the measured drift times from the expected drift times $t(r_{track})$ on the ratio $\frac{q_{ML}}{q}$ shown in Figure~\ref{Fig9} exhibits the predicted linear dependence of the time walk on the ratio $\frac{q_{ML}}{q}$. The profile of the scatter plot in Figure~\ref{Fig9} provides a time-slewing correction curve linear in $\frac{q_{ML}}{q}$ which can be used to correct the measured drift times for their walk with the pulse heights. This time-slewing correction leads to a substantial improvement of the spatial resolution of the drift tubes. The spatial resolution obtained with the time-slewing correction is compared to the spatial resolution without time-slewing corrections in Figure~\ref{Fig10}. The time-slewing correction improves the spatial resolution at all track impact radii, by about 50~$\mu$m at small radii, by about 10~$\mu$m at large radii (reflecting the decrease of the drift velocity with increasing radius).

\begin{figure}[hbt]
\begin{center}
    \includegraphics[width=0.5\linewidth]{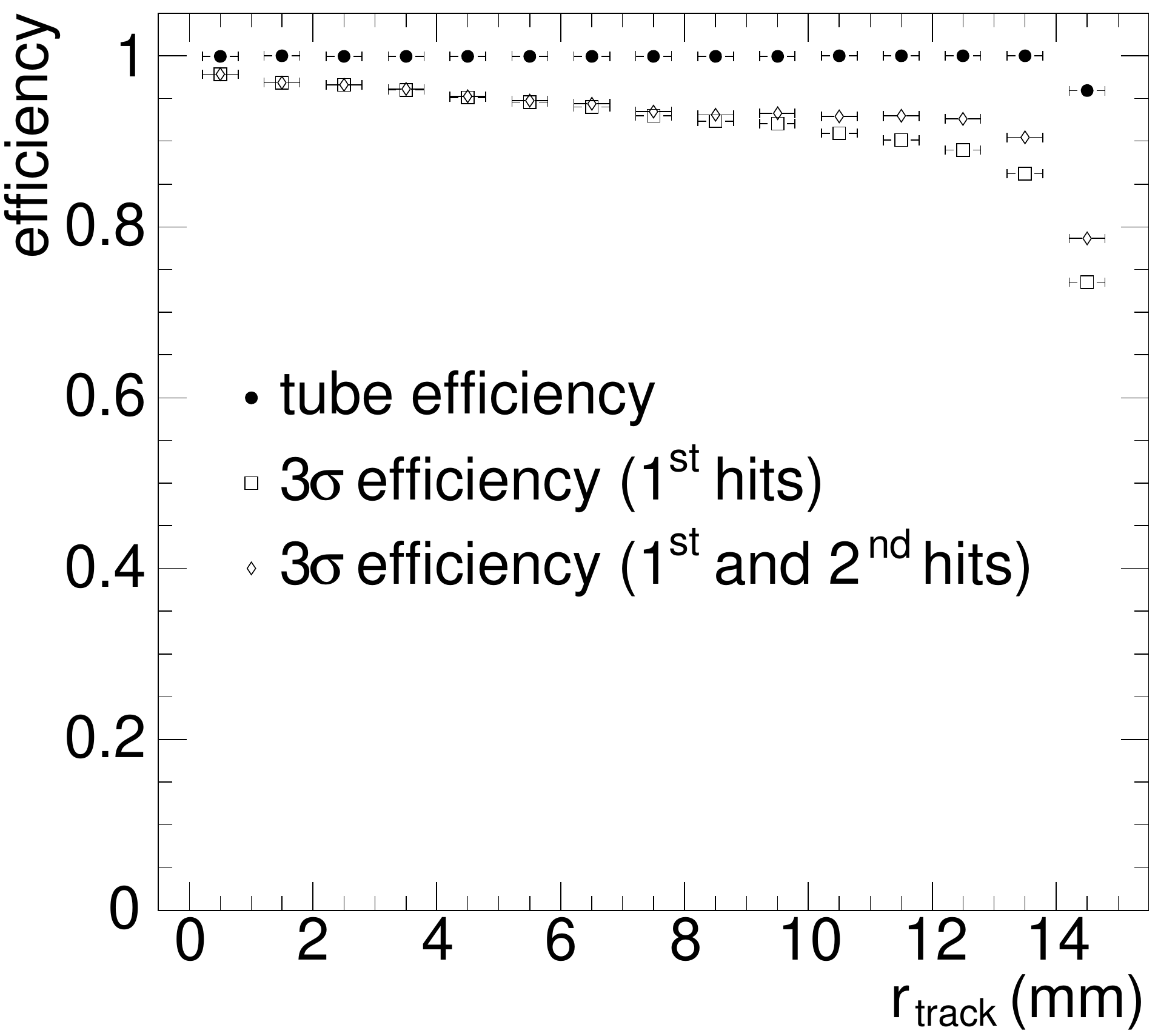}
    \caption{\label{Fig11}Muon detection efficiencies of a tube as a function of the track impact radius.}
\end{center}
\end{figure}

We close this section with the measurement of the muon detection efficiency of the tubes. We distinguish two kinds of efficiencies: (1) The tube efficiency is defined as the probability that a tubes gives a hit when traversed by a muon within the sensitive volume of the tube ($r_{track}<14.6$~mm). (2) The 3$\sigma$~efficiency requires in addition that the drift time associated with the the hit corresponds to a radius $r(t)$ which agrees with the track distance $r_{track}$ from the anode wire of the tube within 3 times the spatial resolution at $r_{track}$. The measurement of both quantities is presented in Figure~\ref{Fig11}. The tube efficiency is 100\% for radii less than 14.2~mm. It drops to 0 from 14.2~mm to 14.6~mm radial distance because of the decreasing number of primary ionization electrons as a consequence of the decreasing muon track length in the gas volume. The overall 3$\sigma$~efficiency is only 93\%. It is lower than the tube efficiency due to the $\delta$~electrons knocked out of the tube wall by the impinging muon and passing the wire at a smaller distance than the muon. The 3$\sigma$~efficiency drops with increasing track impact radius because of the rising probability of a $\delta$~electron hit masking a muon hit. A part of the masked muon hits can be recovered by operating the read-out electronics at the minimum dead time of 200~ns and considering second hits. As the drift time $t(r)\leq200$~ns for $r\leq7.5$~mm (see Figure~\ref{Fig3}), muon hits can only be recovered for track radii greater than 7.5~mm. The overall gain in efficiency by including second hits is less than 1\%. The gain in efficiency will be much larger when the chamber is operated under high $\gamma$~radiation background as will be shown in Section~\ref{Sec43}.

\subsection{\label{Sect3.2}Operation in magnetic fields}
\begin{figure}[hbt]
\begin{center}
    \includegraphics[width=0.5\linewidth]{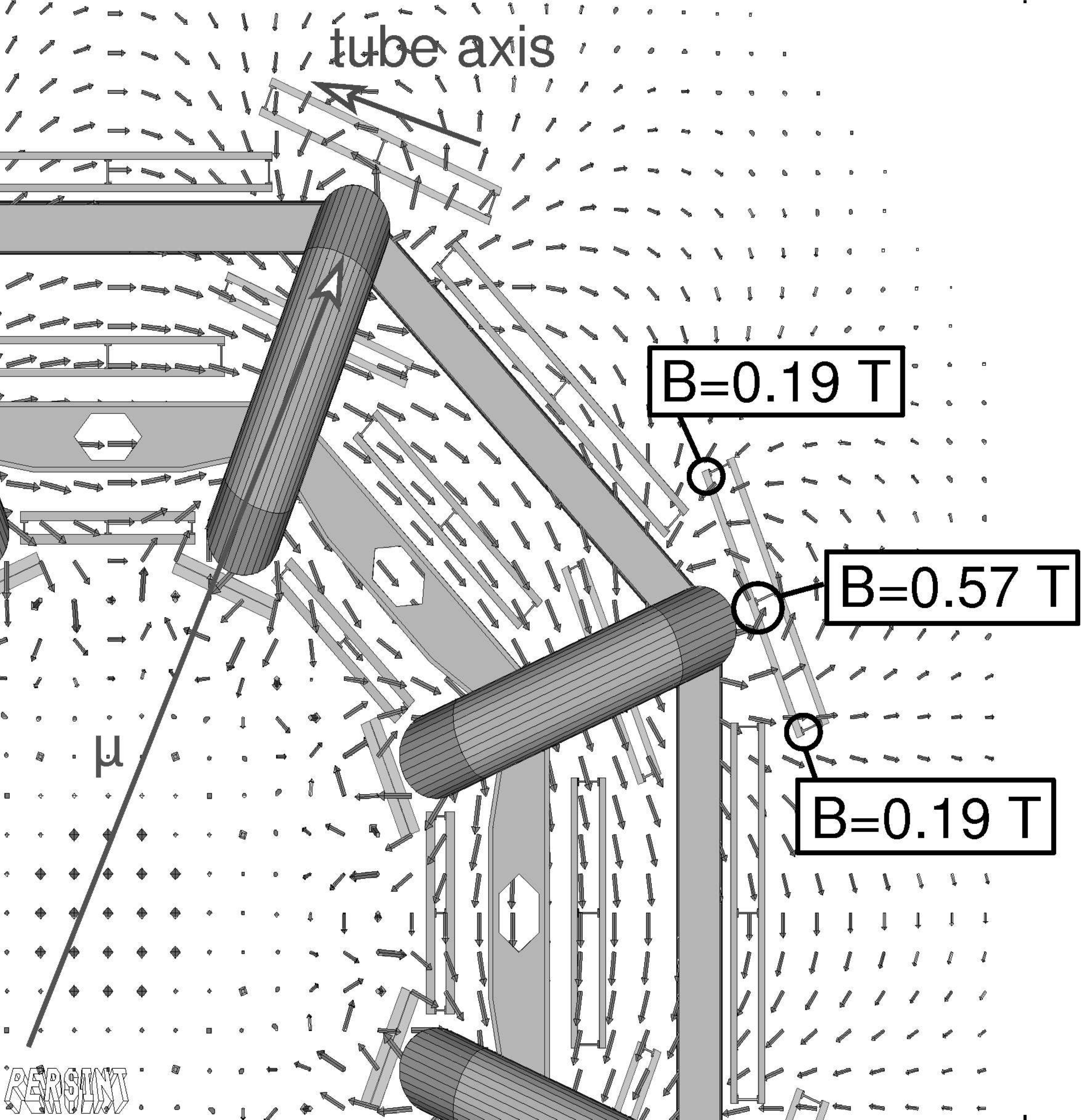}
    \caption{\label{Fig12}Magnetic field configuration in the barrel part of the ATLAS muon spectrometer. The outermost MDT chambers experience spatial field variations of up to 0.4~T.}
\end{center}
\end{figure}

So far we have discussed the operation of the MDT chambers in regions of no magnetic fields as is the case in the end caps of the muon spectrometer. In the barrel part of the muon spectrometer, however, the chambers are operated in a magnetic field of about 0.5~T. Figure~\ref{Fig12} shows a schematic of the field configuration in the barrel region of the muon spectrometer. The magnetic field produced by the superconducting coils is toroidal inside the coils and very non-uniform outside the coils. The MDT chambers are installed with their anode wires aligned parallel to the toroidal field lines in order to measure the magnetic field deflections of the muons emerging from the proton-proton interaction point. The chambers which are mounted on the outer side of a coil see a magnetic field strongly varying in strength and orientation. The magnetic field strength rapidly decreases from one tube layer to the next one in the chambers of the outermost ring. The strength even changes significantly along the anode wires of single tubes. These variations can amount to up to 0.4~T. These chambers have no regions of considerable size with the same magnetic-field configuration which makes a good understanding of the magnetic-field dependence of the space drift-time relationship mandatory.

If a muon traverses the gas volume of a drift tube, it ionizes gas atoms and molecules along its trajectory. The electric field inside the tube pulls the freed electrons towards the anode wire where they trigger the avalanche which gives rise to the measured signal. The shortest drift path of the electrons in the absence of a magnetic field is orthogonal to the muon trajectory and of length $r_{track}$, the distance of the muon trajectory from the anode wire. As the discriminator threshold in the read-out electronics of the tube is set to a low value, the time $t$ measured by the tube is the drift time of the electrons travelling the distance $r_{track}$ from the muon trajectory to the anode wire. The motion of the drifting electrons in the absence of a magnetic field is usually described by the Langevin equation 
\begin{eqnarray}
    \stackrel{..}{x}_2 = -\frac{\stackrel{.}{x}_2}{\tau}
                        +\frac{e}{m}E(x_2)
    \nonumber
\end{eqnarray}
where $\vec{x}_2$ points into the drift direction of the electrons, $E(x_2)=\frac{U_0}{\ln\frac{r_{max}}{r_{min}}}\frac{1}{x_2}$ ($r_{min}:=0.025$~mm, $r_{max}:=14.6$~mm) is the electric field inside the tube, and $\tau$ is the mean time between subsequent collisions of the drifting electrons with a gas molecule. For most gases and also for Ar:CO$_2$, the acceleration $\stackrel{..}{x}_2$ of the electrons is small compared to the electric pull $\frac{e}{m}E(x_2)$ such that the drift velocity equals
\begin{eqnarray}
    \stackrel{.}{x}_2 = -\frac{e}{m}E(x_2) \cdot \tau
    \label{Eqn1}
\end{eqnarray}
to good approximation. The interpretation of Equation (\ref{Eqn1}) is that the electrons acquire their kinetic energy in between the collisions with the gas molecules or gas atoms by which they are slowed down. This interpretation indicates that the time between collisions or, equivalently, the collision rate cannot be the only quantity determining the drift velocity besides the electric field. For assume that there were two different gases in which the electron's drift velocity at a given field were the same, but that in one of the two gases the collisions of the electrons with the gas molecules were purely elastic while in the other they were, at least partially, inelastic. In the second case the electrons would lose more energy in a single collision than in the first case. As a consequence, less collisions lead to the observed drift velocity in the first second case than in the first case. Langevin's picture of an electron experiencing a frictional force $-\frac{\stackrel{.}{x}_2}{\tau}$ in addition to the electric force $\frac{e}{m}E(x_2)$ suggests a simple modification of his equation which accounts for the degree of inelasticity of the electron-molecule collisions. The existence of inelastic collisions can be described by a frictional force which is not proportional to the velocity $\stackrel{.}{x}_2$ of the electron, but to a higher power $\stackrel{.}{x}_2^{1+\epsilon}$ with $\epsilon>0$. From now on we shall consider the modified Lagevin equation
\begin{eqnarray}
    \stackrel{..}{x}_2 = - \left(\frac{\stackrel{.}{x}_2}{\tau_\epsilon}\right)
                                                        ^{1+\epsilon}
                         + \frac{e}{m}E(x_2)
    \label{Eqn2}
\end{eqnarray}
with small $\epsilon\leq0$.

In the next step we consider the presence of a constant magnetic field $\vec{B}$ inside the tube. If the magnetic field is moderate, the deflection of the electrons from a straight trajectory is so small that it is still justified to assume that the measured drift time is the time when $x_2$ equals the radius $r_{min}$ of the anode wire. We decompose $\vec{B}$ into a component $\vec{B}_\parallel$ parallel to the $x_2$~axis and a component $\vec{B}_\perp$ orthogonal to it. We chose the $x_2$ axis as above, $x_1$ parallel to $\vec{B}_\perp$ and $x_3$ perpendicular to $x_1$ and $x_2$. Equation~(\ref{Eqn1}) transforms into the set of equations
\begin{eqnarray}
    \stackrel{..}{x}_1 &=&
        -\left( \frac{\stackrel{.}{x}_1}{\tau_\epsilon}\right)^{1+\epsilon}
        +\frac{e}{m}E_1(\vec{x})
        +\frac{e}{m}\stackrel{.}{x}_1 B_\parallel,
    \nonumber \\
    \stackrel{..}{x}_2 &=&
        -\left( \frac{\stackrel{.}{x}_2}{\tau_\epsilon}\right)^{1+\epsilon}
        +\frac{e}{m}E_2(\vec{x})
        -\frac{e}{m}\stackrel{.}{x}_3 B_\perp,
    \nonumber \\
    \stackrel{..}{x}_3 &=&
        -\left( \frac{\stackrel{.}{x}_3}{\tau_\epsilon}\right)^{1+\epsilon}
        +\frac{e}{m}E_3(\vec{x})
        -\frac{e}{m}(\stackrel{.}{x}_1 B_\parallel
                        - \stackrel{.}{x}_2 B_\perp).
    \nonumber
\end{eqnarray}
These equations simplify due to the assumptions made before:
\begin{enumerate}
    \item   The acceleration is small enough to be negligible, i.e.
            $\stackrel{..}{\vec{x}}=0$.
    \item   In case of moderate magnetic fields the electron drift is mainly
            along the $x_2$ axis. Hence $E_1(\vec{x})\approx0$,
            $E_3(\vec{x})\approx0$, and 
            $E_2(\vec{x})\approx\frac{U_0}{\ln\frac{r_{max}}{r_{min}}}\cdot
            \frac{1}{x_2}=E(x_2)$. As a consequence $\stackrel{.}{x}_1\approx0$
            or -- more precisely -- $|\stackrel{.}{x}_1|\ll|\stackrel{.}{x}_2|$.
\end{enumerate}
The simplifications lead to the system of two coupled equations
\begin{eqnarray}
    0 &=& -\left(\frac{\stackrel{.}{x}_2}{\tau_\epsilon}\right)^{1+\epsilon}
        +\frac{e}{m}E(x_2)
        -\frac{e}{m}\stackrel{.}{x}_3 B_\perp,
    \nonumber\\
    0 &=& -\left(\frac{\stackrel{.}{x}_3}{\tau_\epsilon}\right)^{1+\epsilon}
          +\frac{e}{m}\stackrel{.}{x}_2 B_\perp
    \nonumber
\end{eqnarray}
providing a single equation for the electron drift
\begin{eqnarray}
    0 = -\left(\frac{\stackrel{.}{x}_2}{\tau_\epsilon}\right)^{1+\epsilon}
        +\frac{e}{m}E(x_2)-
        \left(\frac{e}{m}B_\perp\right)^{1+\frac{1}{1+\epsilon}}
        \tau_\epsilon\stackrel{.}{x}_2^{\frac{1}{1+\epsilon}}
    \nonumber
\end{eqnarray}
or, for $\epsilon\ll1$,
\begin{eqnarray}
    0 = -\left(\frac{\stackrel{.}{x}_2}{\tau_\epsilon}\right)^{1+\epsilon}
        +\frac{e}{m}E(x_2)
        -\left(\frac{e}{m}B_\perp\right)^{2-\epsilon}\tau_\epsilon
         \stackrel{.}{x}_2^{1-\epsilon}.
    \nonumber
\end{eqnarray}
So
\begin{eqnarray}
    \stackrel{.}{x}_2
        \left[
        1+\left(\frac{e}{m}B_\perp\right)^{2-\epsilon}\tau_\epsilon^{2+\epsilon}
                \stackrel{.}{x}_2^{-2\epsilon}
        \right]^{1-\epsilon}
        = \left[
            \frac{e}{m}E(x_2)
          \right]^{1-\epsilon}
          \tau_\epsilon
    \nonumber
\end{eqnarray}
where the right hand side is the drift velocity $v_0$ in the absence of the magnetic field according to Equation~(\ref{Eqn2}) in the approximation $\stackrel{..}{x}_2=0$. As shown later, the drift velocity $\stackrel{.}{x}_2$ for $\vec{B}\neq0$ differs from $v_0$ by less than 10\% for magnetic field strengths characteristic for ATLAS. We can, therefore, approximate $\stackrel{.}{x}_2^{-2\epsilon}$ in the parenthesis on the left hand side by $v_0^{-2\epsilon}$ and ignore the $-\epsilon$~term in the exponent of the parenthesis without introducing a systematic error of more than 1\% as long as $\epsilon\lesssim0.1$. By integrating the resulting equation
\begin{eqnarray}
    \frac{1}{\stackrel{.}{x}_2} =
        \frac{1}{v_0}
            \left[
                1+\left(\frac{e}{m}B_\perp\right)^{2-\epsilon}
                    \tau_\epsilon^{2+\epsilon}v_0^{-2\epsilon}
            \right]
    \nonumber
\end{eqnarray}
one obtains the drift time
\begin{eqnarray}
    t(r,\vec{B}) = t(r,\vec{B}=0)
                    +B_\perp^{2-\epsilon}
                        \int\limits_{r_{min}}^{r}
                        \frac{v_0^{1-\epsilon}}{E^{2-\epsilon}(x_2)}\,dx_2.
    \label{Eqn3}
\end{eqnarray}
So the drift time $t(r,\vec{B})$ in the presence of the magnetic field $\vec{B}$
equals the drift time $t(r,0)$ in the absence of the magnetic field plus a term
which factorizes into a part depending only on the magnetic field configuration inside the tube and a part depending only on the drift velocity $v_0$ in the absence of the magnetic field and the electric field inside te tube.

\begin{figure}[hbt]
\begin{center}
    \includegraphics[width=0.5\linewidth]{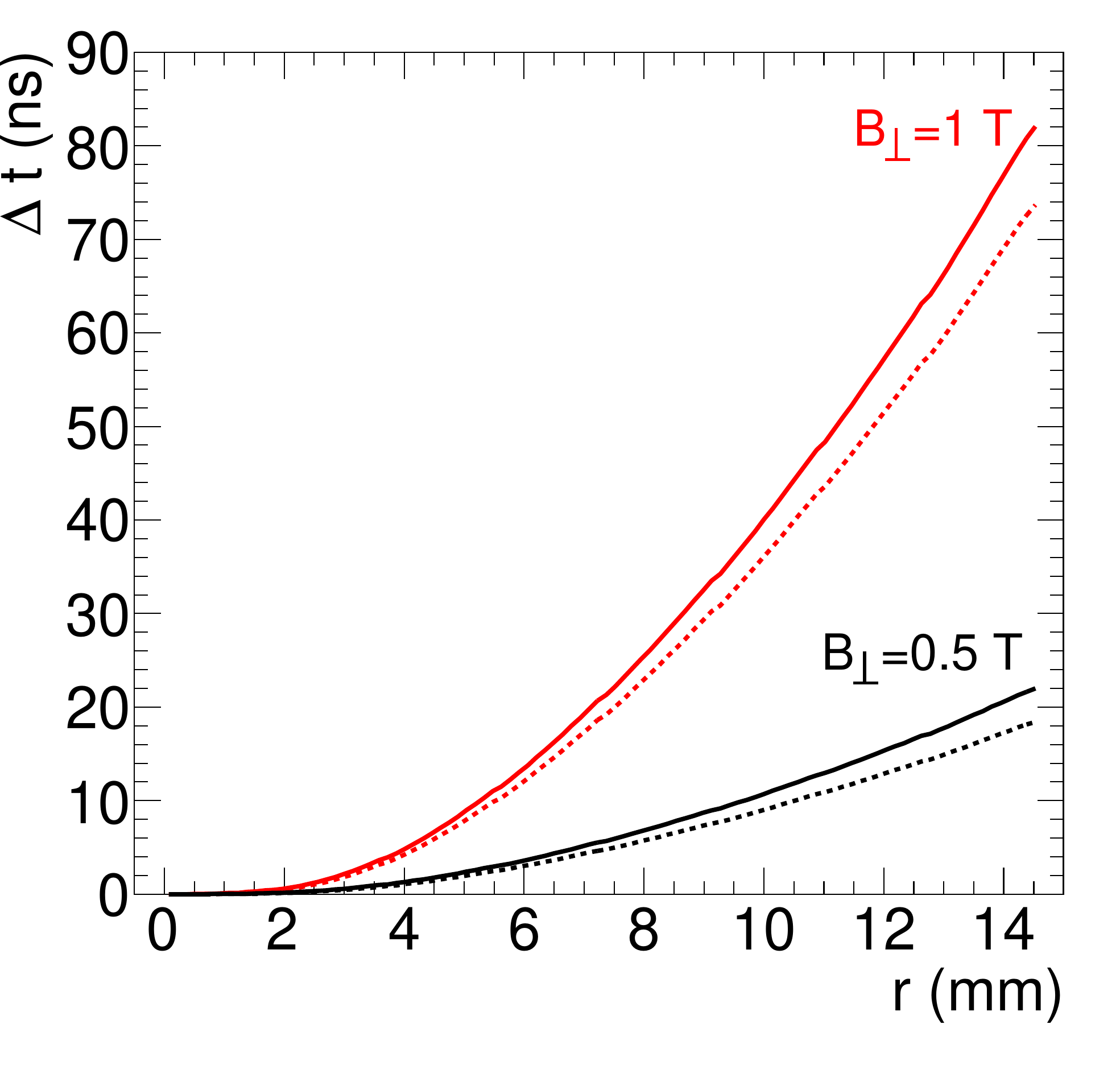}
    \caption{\label{Fig13}$\Delta t(r,\vec{B})$ for $B_\perp=0.4$~T, $1$~T according to Equation~(\ref{Eqn3}). Solid line: $\epsilon=0.1$. Dashed line: $\epsilon=0$.}
\end{center}
\end{figure}

The change of the $t$-$r$ relationship $\Delta t(r,\vec{B}):=t(r,\vec{B})-t(r,0)$ due to a nonvanishing field $B_\perp$ is illustrated in Figure~\ref{Fig13} for $B_\perp=0.5$~T, the typical field strength in the barrel MDT chambers, and $B_\perp=1$~T for two values of $\epsilon$: $\epsilon=0$ (purely elastic electron-molecule collisions) and $\epsilon=0.1$ (partially inelastic collisions). $\Delta t(r, \vec{B})$ is small for small radii and rises rapidly with increasing drift radius. At the characteristic field strength of 0.5~T in the ATLAS muon spectrometer, the maximum drift time is increased by about 20~ns from its value for vanishing magnetic field. The change of the maximum drift time is 2~times bigger for $B_\perp=1$~T due to the $B_\perp^{2-\epsilon}$ scale factor in the formula for $\Delta t(r,\vec{B})$. $\Delta t(r,\vec{B})$ is larger in case of partially inelastic collisions than in case of purely elastic collisions between the drifting electrons and the gas molecules. This behaviour is caused by the fact that, for the same drift velocity at $\vec{B}=0$, the mean time between electron-molecule collisions is greater in the partially inelastic than the purely elastic case as discussed above. As the deflection of the electrons is proportional to the time between collisions times the Lorentz force, the electrons arrive at the anode wire with larger time delay in the presence of inelastic collisions than in the purely elastic case.

\begin{figure}[hbt]
\begin{center}
    \includegraphics[width=0.5\linewidth]{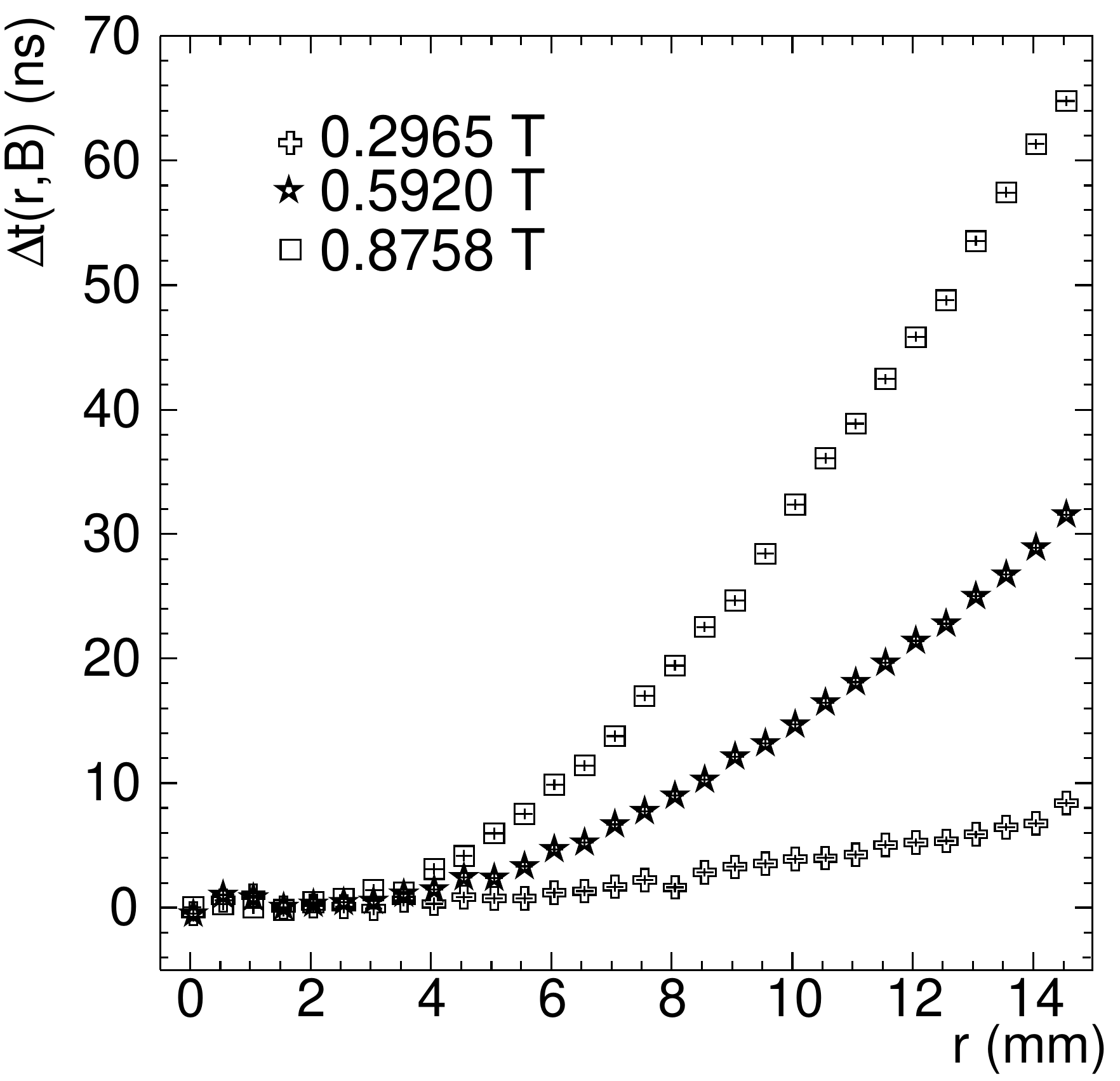}
    \caption{\label{Fig14}Test-beam measurement of $\Delta t(r,\vec{B})$ for three different field strengths with the magnetic field $\vec{B}$ aligned with the anode wires of the tubes.}
\end{center}
\end{figure}

Figure~\ref{Fig14} summarizes the time changes $\Delta t(r,\vec{B})$ measured in the test beam for three field strengths with the magnetic field direction parallel to the anode wires of the tubes. The graphs have the same shapes as the predictions shown in Figure~\ref{Fig13}. The error bars of $\Delta t$ in Figure~\ref{Fig14} are the sum of the statistical errors of the drift-time measurements which amount to $\sqrt{2}\cdot 0.2$~ns=$0.3$~ns, and the systematic uncertainty of $\Delta t$, which is of the same order of magnitude. According to Equation~(\ref{Eqn3}) there are three potential sources of a systematic uncertainty of $\Delta t(r, \vec{B})$: (1) the limited knowledge of the magnetic field strength $B$, (2) temperature fluctuations affecting the drift velocity $v_0$, and (3) an uncertainty in the absolute value of the operating voltage $U_0$ influencing the electric field inside the tubes. As the magnetic field was measured by a Hall probe attached to the small drift-tube chamber with 1~mT accuracy, the systematic error arising from the limited knowledge of the magnetic field is less than 0.1~ns, hence negligible compared to the statistical error of 0.3~ns. The gas temperature fluctutations were measured to be less than 0.2~K leading to negligible fluctuations of the space drift-time relationships. According to the data sheet of the power supply used in the test-beam measurements, the scale of the operating voltage is correct to 1\% resulting in a non-negligible 1\% uncertainty of $\Delta t(r,\vec{B})$ reaching the statistical error of $\Delta t(r,\vec{B})$.

\begin{figure}[hbt]
\begin{center}
    \includegraphics[width=0.55\linewidth]{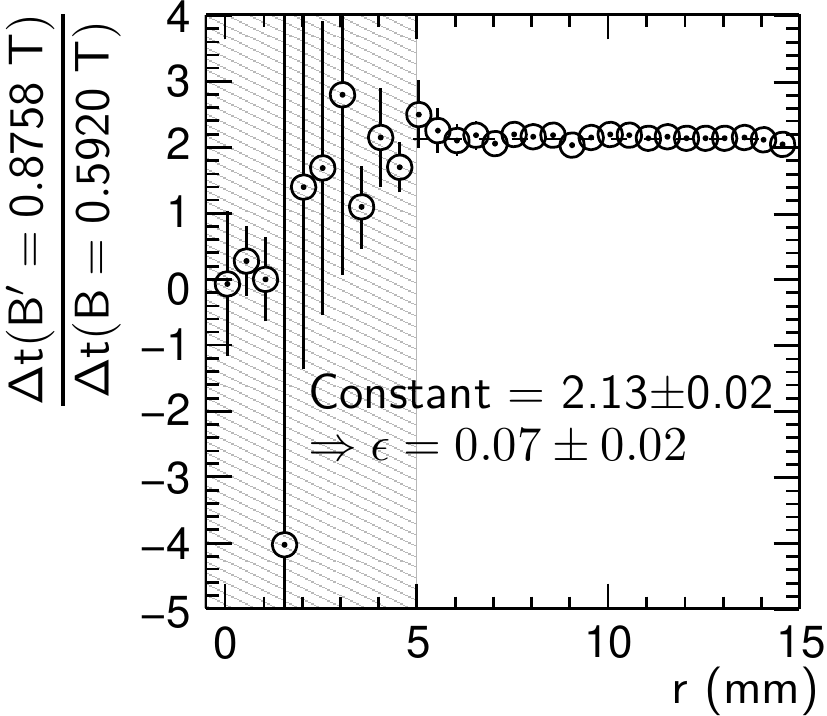}
    \caption{\label{Fig15}Measured ratio $\frac{\Delta t(r,B'_\perp=0.8758~\mathrm{T})}{\Delta t(r,B_\perp=0.5920~\mathrm{T})}$. Error bars contain statistical and systematic uncertainties. The shaded area is excluded from the fit because there $\Delta t(r, B>0.5)=0$ within errors.}
\end{center}
\end{figure}

Equation~\ref{Eqn3} predicts a factorization of $\Delta t(r, \vec{B})$ into a purely magnetic field dependent part and a purely magnetic field independent part. So the ratio $\frac{\Delta t(r,\vec{B}')}{\Delta t(r,\vec{B})}$ should be a constant equal to $\left(\frac{B'_\perp}{B_\perp}\right)^{2-\epsilon}$. The measured ratio $\frac{\Delta t(r,\vec{B}')}{\Delta t(r,\vec{B})}$ for $B'_\perp=0.8758$~T and $B_\perp=0.5920$~T presented in Figure~\ref{Fig15} as a function of the drift radius $r$ is compatible with a constant of value $2.13\pm0.02$ corresponding to $\epsilon=0.07\pm0.02$. $\epsilon>0$ indicates that inelastic collisions of electrons with CO$_2$ molecules (excitation of rotational and vibrational states of CO$_2$) play a non-negligible role in the electron drift in the Ar:CO$_2$ gas mixture of the ATLAS drift tubes.

\begin{figure}[hbt]
\begin{center}
    \includegraphics[width=0.5\linewidth]{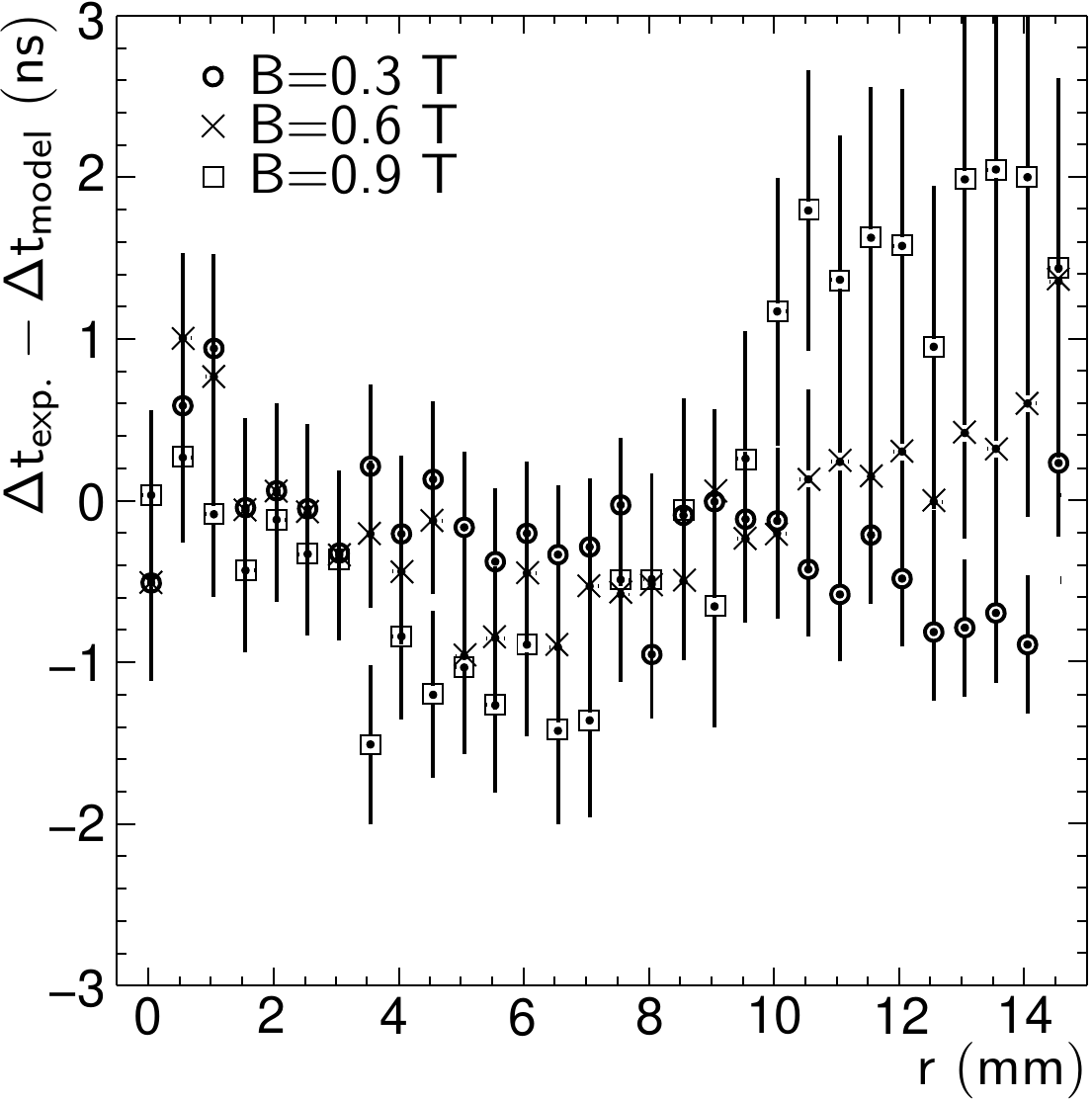}
    \caption{\label{Fig16}Residuals of the fit of the model for $\Delta t_{model}(r,\vec{B}, \epsilon)$ to the measured drift-time changes $\Delta t_{exp}(r,\vec{B})$ of Figure~\ref{Fig14}. Error bars contain statistical and systematic uncertainties.}
\end{center}
\end{figure}

A more precise value of the inelasticity parameter $\epsilon$ is obtained by fitting the expected drift-time changes $\Delta t(r, \vec{B}, \epsilon)$ simultaneously to the measurements presented in Figure~\ref{Fig14}. The common fit gives $\epsilon=0.101\pm0.005$. The residuals $\Delta t(r,\vec{B})-\Delta t_{model}(r, \vec{B}, \epsilon=0.101)$ in Figure~\ref{Fig16} show that the model describes the measurements for $B_\perp=0.3$~T and $B_\perp=0.6$~T -- field strength characteristic for the ATLAS muon spectrometer -- while it systematically deviates by up to 1~ns from the experimental data points for $B_\perp=0.9$~T. This deviation is caused by the 1\% systematic error in the derived formula as explained above. Yet, the impact of an error of 1~ns on $r(t)$ is less than 20~$\mu$m, so the model is adequate to describe the magnetic field dependence of the space drift-time relationship for the magnetic field strengths encountered in the ATLAS muon spectrometer.

\begin{figure}[hbt]
\begin{center}
    \includegraphics[width=0.5\linewidth]{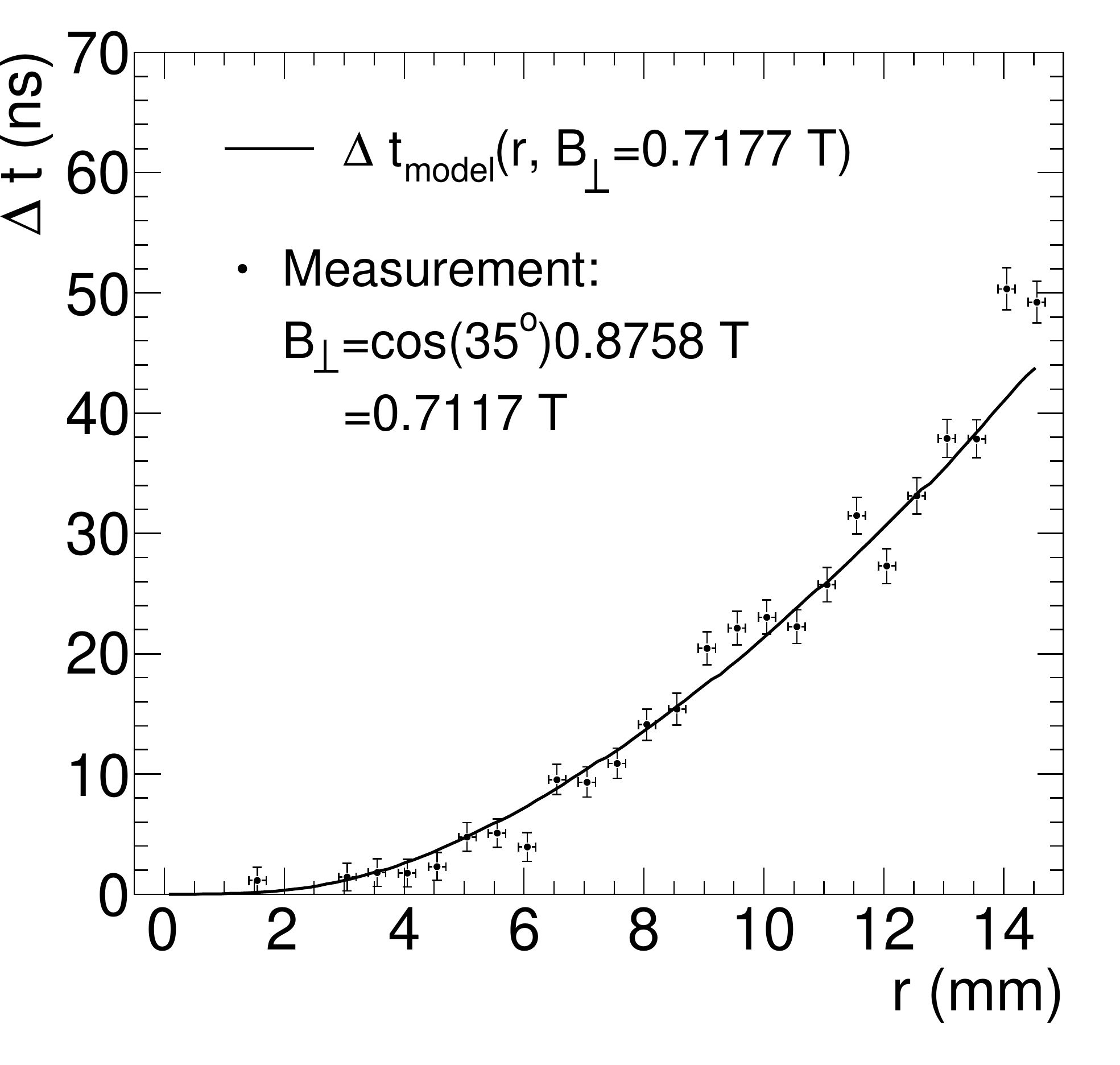}
    \caption{\label{Fig17}Comparison of the change of the drift time $t$ at a given radius $r$ caused by a magnetic field of 0.8758~T strength at an angle of $35^o$ with respect to the tube axis and orthogonal to the muon beam.}
\end{center}
\end{figure}

The rotation of the small drift-tube chamber around the veritcal axis of the set-up leaves the magnetic field vector orthogonal to the main drift direction of the electrons. $B_\perp$ is unaltered by this rotation and, as a consequence, $\Delta t(r,\vec{B})$ is expected and measured to be the same as for $\vec{B}$ parallel to the anode wires. A roation of the chamber around the beam axis by an angle $\beta$, however, changes $B_\perp$ from $B$ to $B\cos\beta$. This leads to a 19\% reduction of $B_\perp$ for $\beta=35^o$ (the largest angle accessible in the test-beam set-up) compared to $\beta=0$. As $\Delta t(r,\vec{B})$ is proportional to $B_\perp^{1.9}$, $\Delta t(r, B, \beta=35^o)=0.32\Delta t(r, B, \beta=0)$. This reduction is observed in Figure~\ref{Fig17} which compares the measurement of $\Delta t(r, B=0.8758~\mathrm{T}, \beta=35^o)$ with the model prediction for this quantity. The data points are compatible with the model prediction. The lare errors and fluctuations of the data points around the model prediction are caused by the limited accuracy of the measurement of the muon position along the wire due to only one measurement plane for this coordinate in the beam hodoscope as
mentioned in Section~\ref{setup}.

\begin{figure}[hbt]
\begin{center}
    \includegraphics[width=0.5\linewidth]{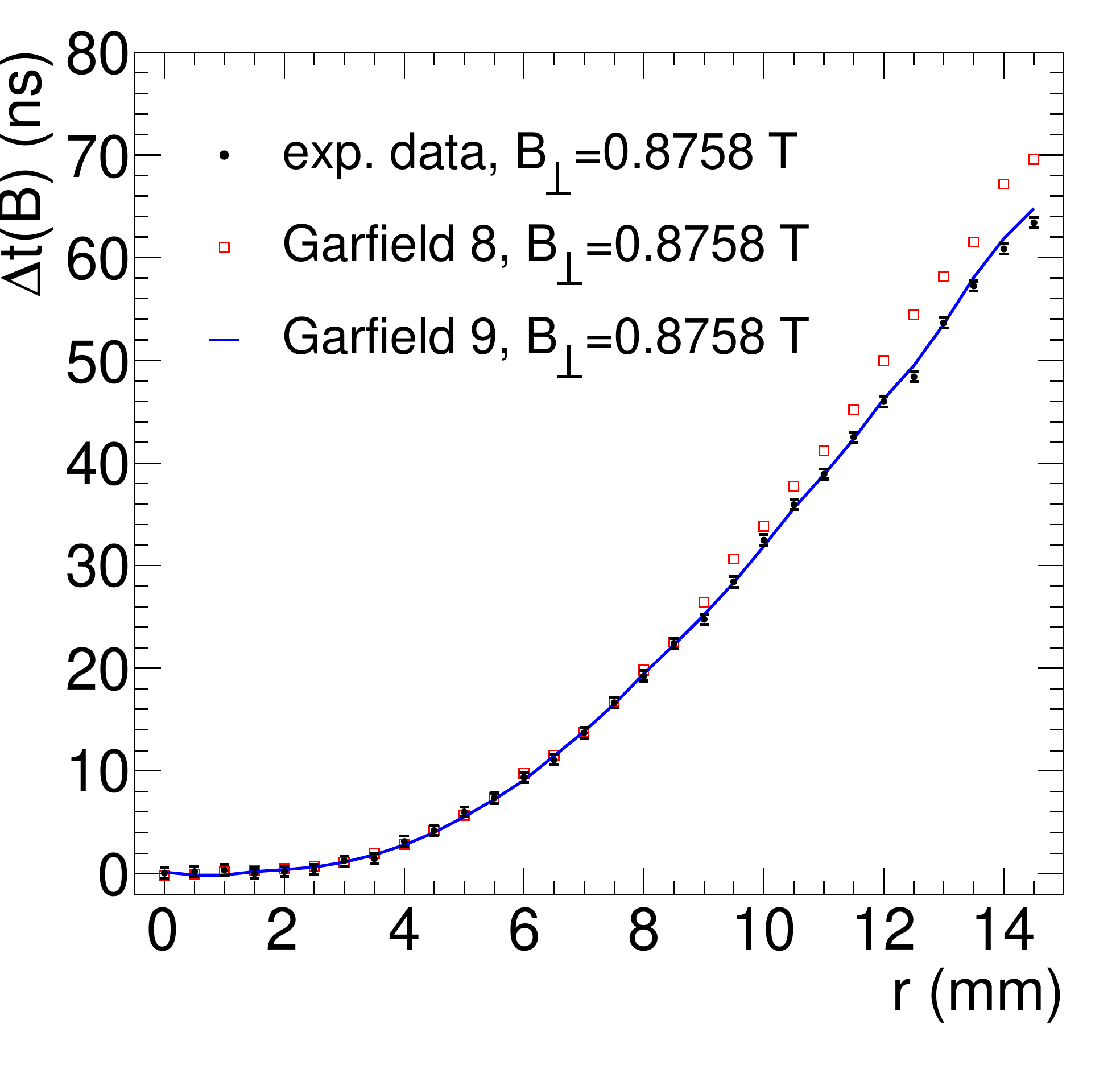}
    \caption{\label{Fig18}Comparison of the Garfield predictions of the magnetic field induced change of $t(r)$ with the experimental data. The Garfield-8 simulation was run with 7.08\% CO$_2$ content in order to reproduce the measured $r$-$t$~relationship without magnetic field. This is no required for Garfield-9 (see Section~\ref{sec31}).}
\end{center}
\end{figure}

We close our studies of drift-tube properties in the absence of radiation background by a comparison of the measured magnetic-field dependence of the space drift-time relationship with predictions of the Garfield simulation programme. It is shown in Section~\ref{sec31} that Garfield version 9 reproduces the measurements of the drift-tube properties in the absence of magnetic field while Garfield version 8 does not. Garfield version 8 requires an increase of the CO$_2$ content from its correct value of 7\% to 7.08\% to predict the measured space drift-time relationship without magnetic field. The predictions of both Garfield versions for the magnetic field induced change of the space drift-time relationship are compared with the measured change for $B_\perp=0.8758$~T in Figure~\ref{Fig18}. The Garfield-8 calculations are performed with 7.08\% CO$_2$ content. Figure~\ref{Fig18} confirms the excellent predictive power of Garfield version 9 which includes excitations of CO$_2$ molecules correctly and shows that Garfield version 8 overestimates $\Delta t(r, B_\perp=0.8758~\mathrm{T})$ significantly. The less predictive power of Garfield-8 is caused by the lack of rotational excitations of CO$_2$ molecules. The accuracy of the Garfield-9 prescition is of the same order as the accuracy of our phenomenological model.

\section{\label{Sec4}Performance of MDT chambers in the presence of high $\gamma$ radiation background}

The ATLAS MDT chambers are operated in a high background of neutrons and $\gamma$~rays at the LHC. The background is particularly high in the end caps of the muon spectrometer with peak values of 100-500~Hz\,cm$^{-2}$ counting rates in the innermost chambers close to the LHC beam pipe at design luminosity of $10^{34}$~cm$^{-2}$s$^{-1}$. At these background rates there is high probability that a muon hits a drift tube when the ions created near the anode wire in the avalanche from a background hit are still drifting to the tube wall. The ions alter the electric field inside the drift tube with several implications discussed below. The average change of the electric field can be calculated in an approximate static model proposed in \cite{Riegler_1, Aleksa_thesis}.

In the static model a cylindrically symmetrical charge distribution is expressed as the sum of the line charge density $\rho_{wire}$ on the anode wire and the space charge distribution $\rho_{ion}$ of the ions. The drift velocity of the ions is given by $\frac{dr}{dt}=\mu E(r)$ with constant mobility $\mu=0.51$~cm$^{2}$(Vs)$^{-1}$. A $\gamma$ background hit deposits on average $Q=1.0\cdot10^{-16}$~C charge in the tube which is multiplied to $G\cdot Q$ at the anode wire with the gain $G$. The total ion charge per time per length of tube is $N_{c}GQ$ at the counting rate $N_c$ per tube length which translates into a ion charge density of
\begin{eqnarray}
    \rho_{ion} = \frac{N_cGQ\,dt}{2\pi r\, dr} = \frac{N_cGQ}{2\pi r\mu E(r)}.
    \nonumber
\end{eqnarray}
The electric field 
\begin{eqnarray}
    E(r) = \sqrt{c_1}\frac{k}{r}\sqrt{1+\frac{r^2}{k^2}},\ 
    c_1:=\frac{N_cGQ}{2\pi\mu\epsilon_0},
    \nonumber
\end{eqnarray}
is the solution of Maxwell's first equation (Gau\ss' law) in cylindrical coordinates
\begin{eqnarray}
    (div\vec{E})(r) = \frac{1}{r}E(r) +
                        \left( \frac{\partial E}{\partial r} \right)(r)
                    = \frac{1}{\epsilon_0}
                        \left[
                        \rho_{wire}\delta(r) + \rho_{ion}(r)
                        \right].
    \nonumber
\end{eqnarray}
The boundary condition that the potential difference between the anode wire and the tube wall equals the operating voltage $U_0$ leads to the transcendental equation for $k$
\begin{eqnarray}
    k\ln\frac{r_{max}(k^2+k\sqrt{k^2+r_{min}^2})}
                        {r_{min}(k^2+k\sqrt{k^2+r_{max}^2})}
        +\sqrt{k^2+r_{max}^2}-\sqrt{k^2+r_{min}^2} =
                U_0\sqrt{\frac{2\pi\mu\epsilon_0}{N_cGQ}}
    \nonumber
\end{eqnarray}
to be solved numerically. The gain $G$ depends on the electric field on the anode wire by Diethorn's formula
\begin{eqnarray}
    G= \left[
       \frac{E(r_{min})}{2.96 E_{min}}
       \right]^{\frac{r_{min}E(r_{min})\ln2}{\Delta V}}
    \nonumber
\end{eqnarray}
with $E_{min}=23$~kV\,cm$^{-1}$ and $\Delta V=34$~V \cite{Riegler_1, Aleksa_thesis}.

\begin{figure}[hbt]
\begin{center}
    \includegraphics[width=0.5\linewidth]{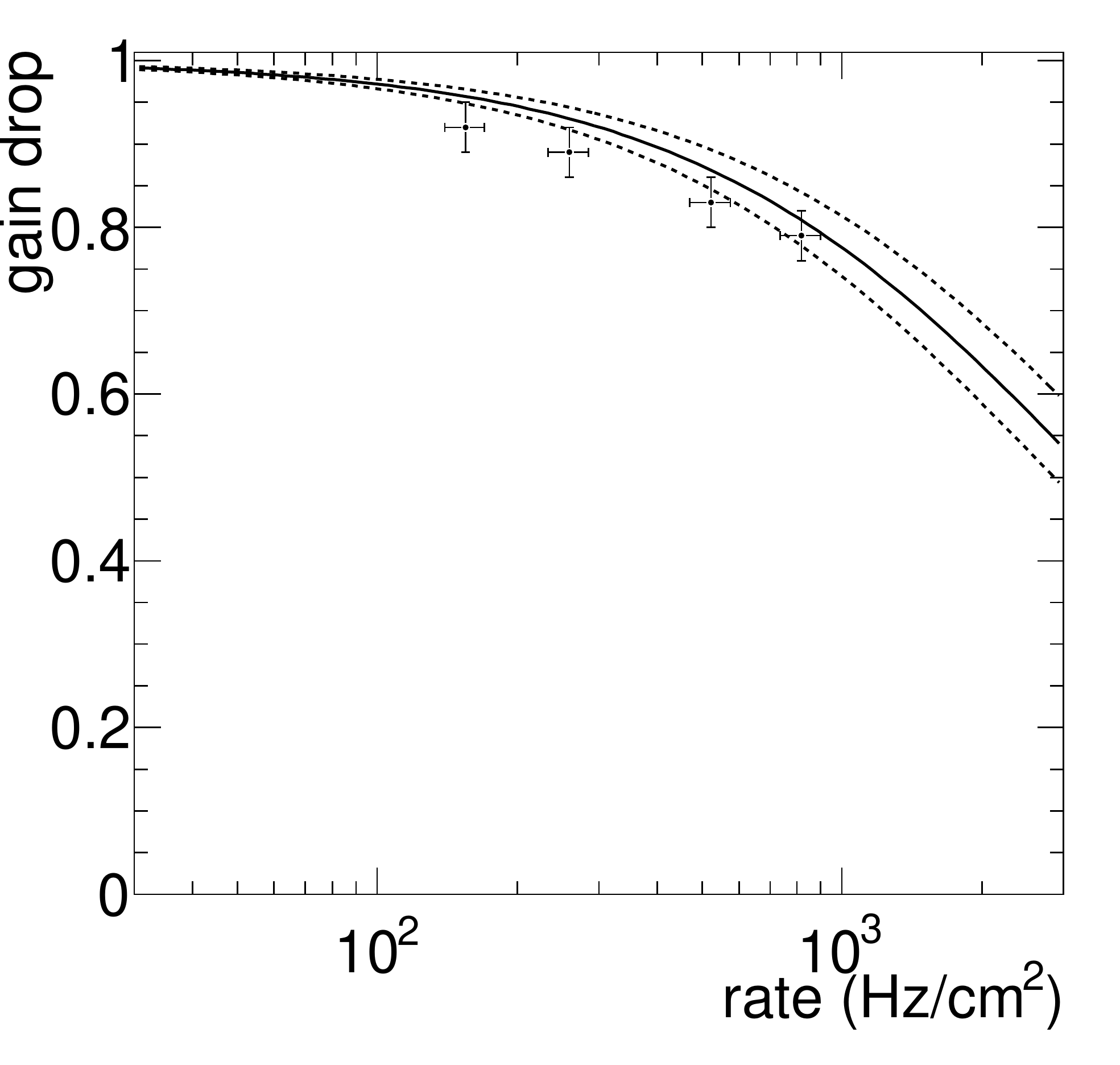}
    \caption{\label{Fig19}Comparison of the measured gain drops with the prediction (solid line) published in \cite{Riegler_1, Aleksa_thesis}. The dashed lines indicate the uncertainty of the prediction.}
\end{center}
\end{figure}

The presence of the ions from background hits reduces the electric field on the anode wire and, as a consequence, the gas gain. Figure~\ref{Fig19} compares the gain drops measured with the ADC of the read-out electronics at different background rates with the prediction of Diethorn's formula including its uncertainties quoted in \cite{Riegler_1, Aleksa_thesis}. The measured gain drop is consistent with the prediction within errors favouring smaller values than predicted.

\subsection{\label{Sec41}Rate dependence of the space drift-time relationship}
The drift velocity of the electrons is a function of the electric field inside the drift tube and can be written as
\begin{eqnarray}
    v_0(r) = \mu_e(E(r))E(r)
    \nonumber
\end{eqnarray}
with the electron mobility $\mu_e$ and $\epsilon=0$. The ion space charge from the background hits changes the electric field from $E_0(r)$ without background radiation to $E_0(r)+\delta E_0(r)$. So the drift velocity changes to
\begin{eqnarray}
    v(r) = \mu_e(E_0(r)+\delta E_0(r))\cdot (E_0(r)+\delta E_0(r)).
    \nonumber
\end{eqnarray}
As $\mu_e(E)$ is approximately proportional to $\frac{1}{E}$ for $r\lesssim5$mm, the drift velocity is independent of the background counting rate for $r\lesssim5$~mm. At larger radii $\mu_e(E)$ approaches a constant $\bar{\mu}_e$
which makes $v(r)$ sensitive to changes of the electric field:
$v(r)\approx v_0(r)+\bar{\mu}_e\cdot\delta E_0(r)$. If $\epsilon$ is small, but nonvanishing, $v(r)\approx v_0(r)+(1-\epsilon)\cdot\bar{\mu}_e\cdot\delta E_0(r)$. The dependence of the drift velocity on the background irradiation leads to a dependence of the space drift-time relationship on the irradiation.

\begin{figure}[hbt]
\begin{center}
    \includegraphics[width=0.5\linewidth]{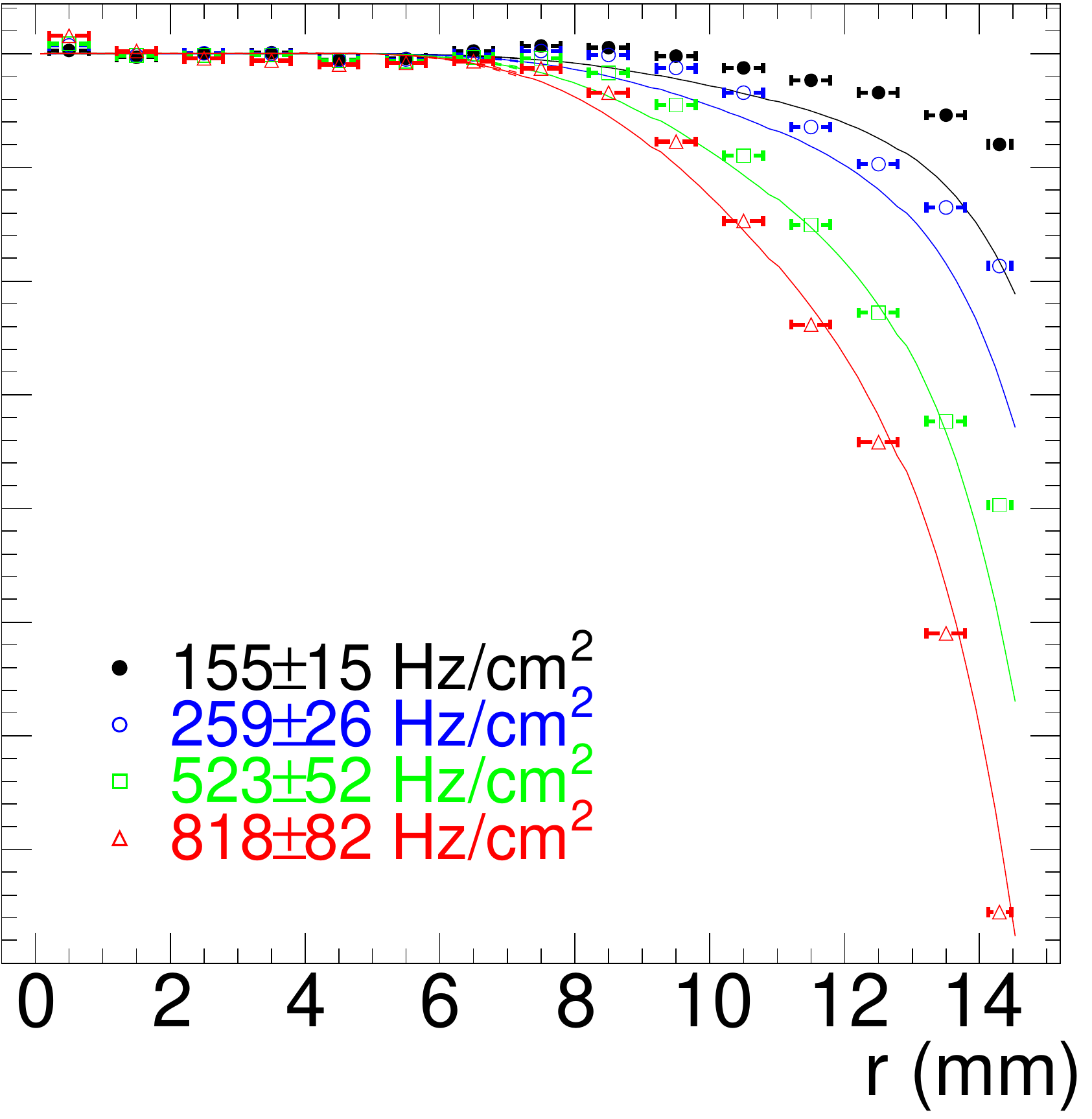}
    \caption{\label{Fig20}Rate dependence of the space drift-time relationship. The points are test-beam measurements. The solid lines represent the predictions of the simple static model.}
\end{center}
\end{figure}

Figure~\ref{Fig20} presents the change $\Delta t(r)$ of $t(r)$ for different background counting rates. $\Delta t(r)=0$ for small radii as $v(r)$ is independent of the irradiation for small radii and drops at larger radii. At a background counting rate of 155~Hz\,cm$^{-2}$ the maximum drift time is reduced by only 4~ns. At the maximum expected counting rate of 500~Hz\,cm$^{-2}$ in the ATLAS muon spectrometer, the change of the maximum drift time is much larger and about 20~ns. The measured changes $\Delta t(r)$ are compared with the prediction of the simple static model. The predictions agree with the measurements within errors for rates above 500~Hz\,cm$^{-2}$, but overestimate the rate dependence of $t(r)$ by up to 5~ns at lower rates. The simple model describes the behaviour of $\Delta t(r)$ qualitatively very well, but, other than the magnetic-field correction derived in Section~\ref{Sect3.2}, it is insufficient for a rate correction of the space drift-time relationship. Since the background radiation will be much more uniform over an MDT chamber than the magnetic field, an accurate rate correction is not needed for the $r$-$t$ calibration and will therefore not be worked out in this article.

\subsection{\label{Sec42}Rate depenence of the spatial resolution}

\begin{figure}[hbt]
\begin{center}
   \includegraphics[width=0.5\linewidth]{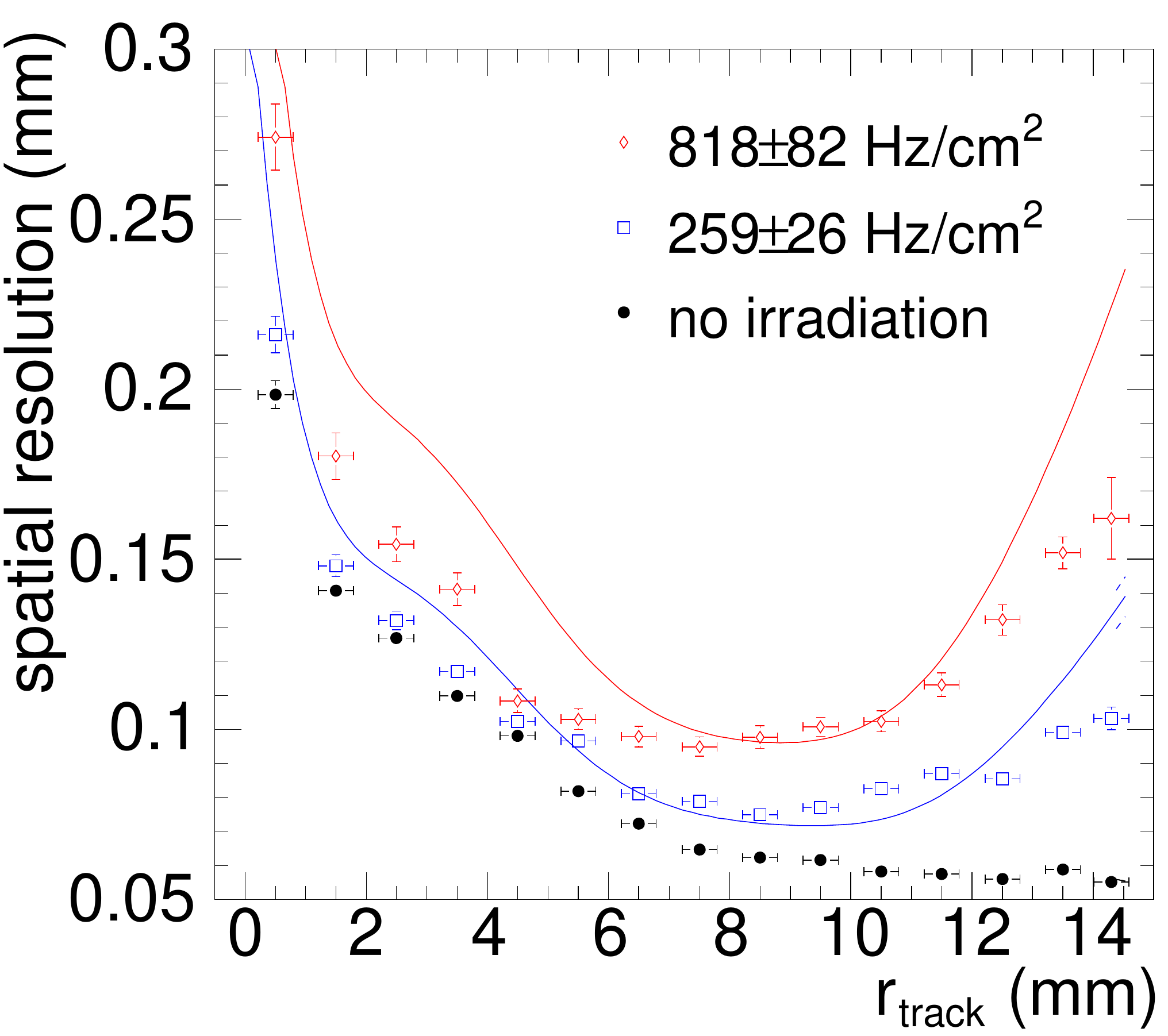}
    \caption{\label{Fig21}Measured single-tube resolution as a function of the track impact radius for different background counting rates. The solid lines show predictions taking into account the influence of the gain drop and space-charge fluctuations on the spatial resolution.}
\end{center}
\end{figure}

The gain drop caused by the radiation background leads to a deterioration of the spatial resolution by a factor equal to the square of the gain drop (as can be derived assuming a linearly rising signal pulse like in Section~\ref{sec31}). An additional source of a resolution degradation is the fluctuation of the ion space-charge density resulting in fluctuations of the electric field inside the tube. The fluctuation of the space drift-time relationship at large radii as a consequence of the field fluctuations deteriorates the spatial resolution further. The measurement of the spatial resolution curves without time-slewing corrections for three different background conditions is shown in Figure~\ref{Fig21}. The resolution degradation for hits close to the wire scales like the square of the gain drop. The space-charge fluctuations lead to an increasing resolution deterioration with increasing track impact radius. The predictions for the resolution presented in Figure~\ref{Fig21} are the quadratic sum of the spatial resolution caused by the gain drop and the expected $r$-$t$~fluctuations due to ion space-charge fluctuations. As in \cite{Riegler_1, Aleksa_thesis} it is assumed that the ion space charge density fluctuations around its mean value $\bar{\rho}_{ion}$ by $\pm\frac{\bar{\rho}_{ion}}{N_c\cdot 1~\mathrm{cm}}$. The prediction is reasonable for a background counting rate of 259~Hz\,cm$^{-2}$, but overestimates the resolution degradation for a rate of 818~Hz\,cm$^{-2}$. This behaviour of the model has also been reported in \cite{Riegler_1, Aleksa_thesis}. It is related to the implicit assumption that a photon conversion does not happen when the ions from a previous conversion are still present in the tube. At a rate of 818~Hz\,cm$^{-2}$ this assumption becomes wrong and the charge fluctuation is no longer Poissonian.

\begin{figure}[hbt]
\begin{center}
    \includegraphics[width=0.49\linewidth]{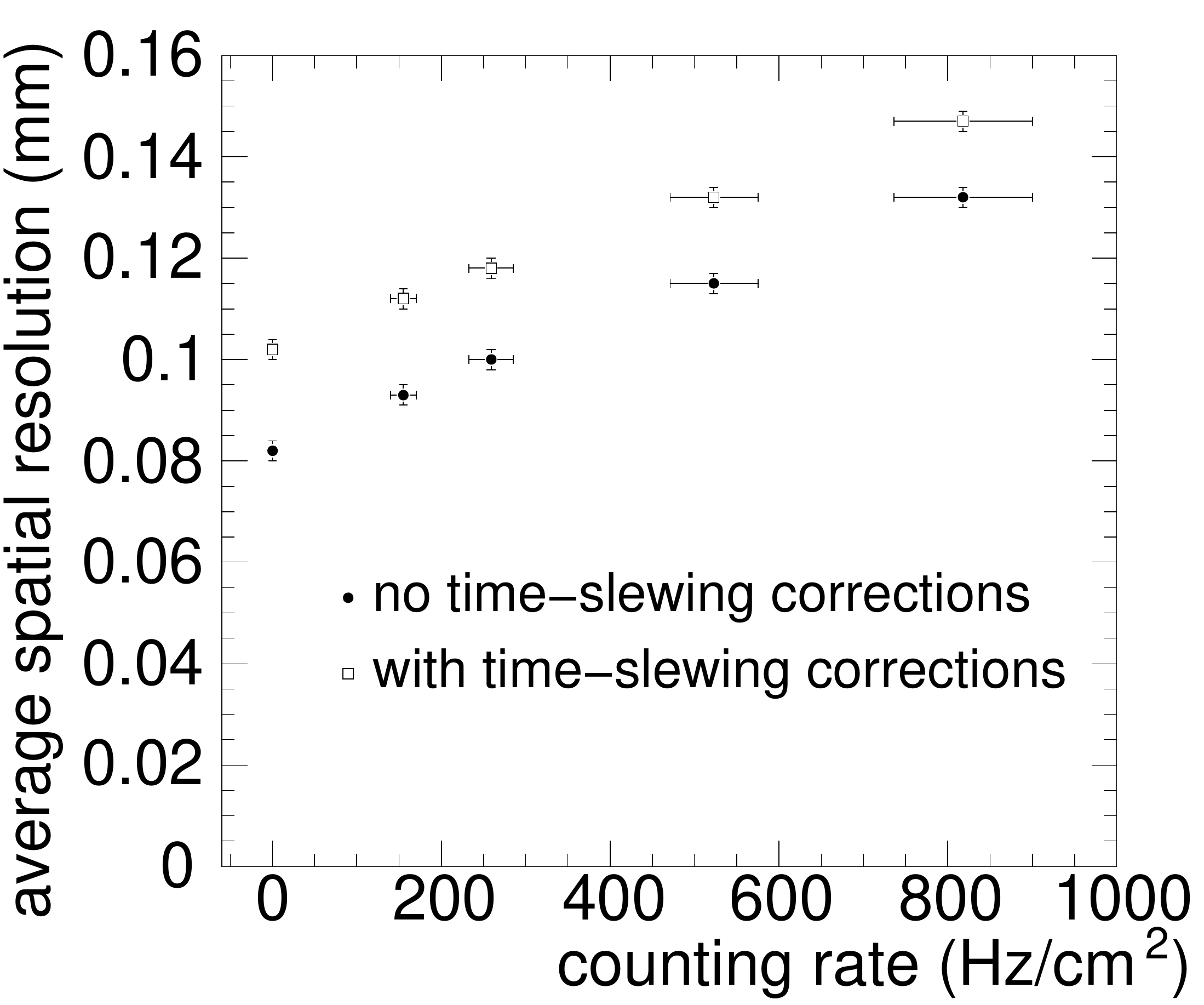}%
    \includegraphics[width=0.49\linewidth]{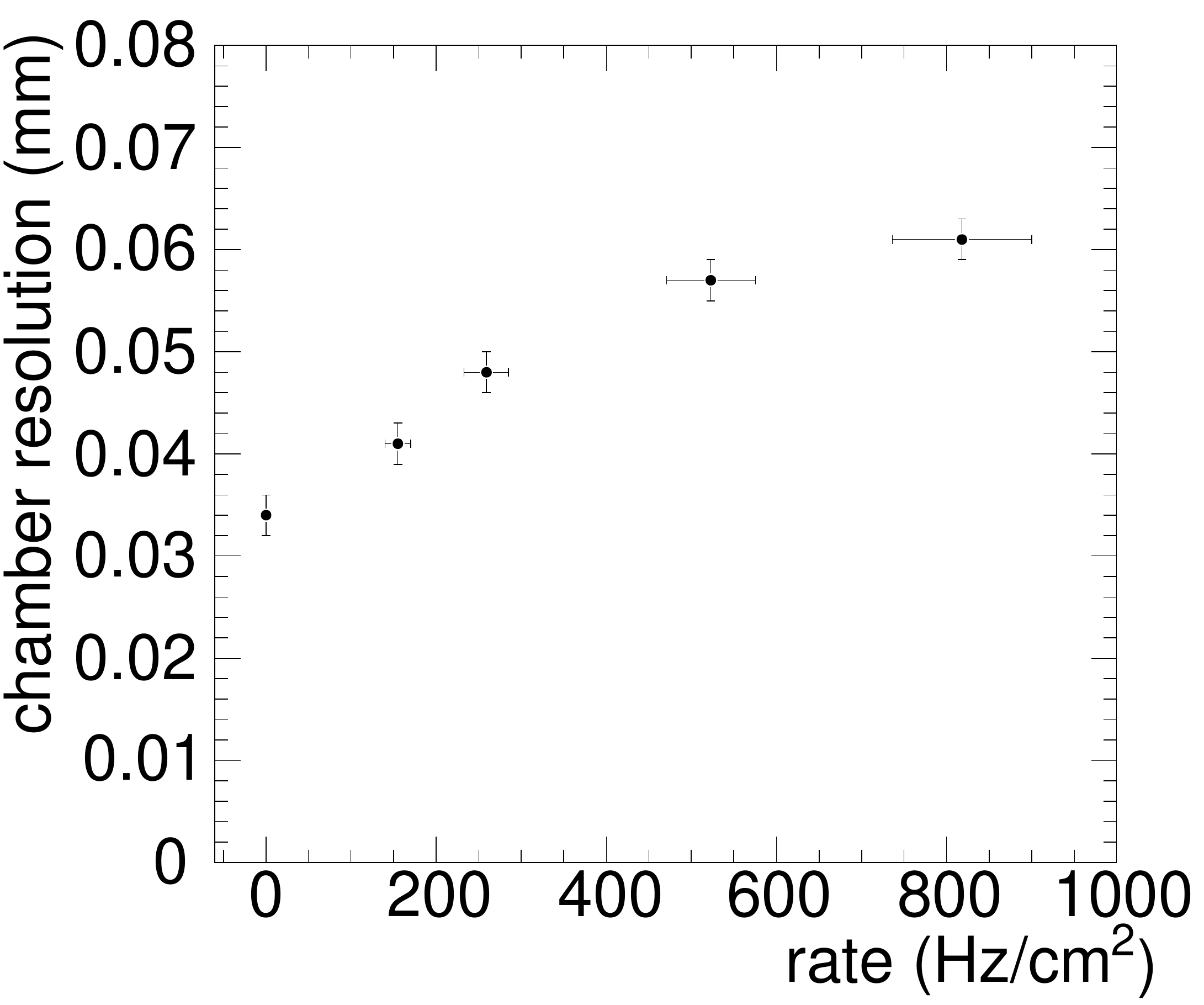}
    \caption{\label{Fig22}Left: Average spatial resolution of a drift tube with and without time-slewing corrections as a function of the background counting rate. Right: Spatial resolution of a six-layer MDT chamber with time-slewing correction as a function of the background couting rate.}
\end{center}
\end{figure}

The spatial resolution of the tubes can be improved by time-slewing corrections as illustrated in Figure~\ref{Fig22} where the average single tube resolution with and without time-slewing corrections is plotted versus the background counting rate. The average spatial resolution is improved by 20~$\mu$m at all rates by the time-slewing correction and kept below 110~$\mu$m up to 500~Hz\,cm$^{-2}$, the highest expected rate in the ATLAS muon spectrometer.
Figure~\ref{Fig22} also shows the time-slewing corrected spatial resolution of a six-layer MDT chamber for perpendicular incidence of the muon beam. The spatial resolution of the chamber is defined as the standard deviation of the track positions middle of the chamber as reconstructed with 6~tube hits around the same track position as measured by the beam hodoscope. The chamber resolution is 35~$\mu$m at low background rates and degrades to about 55~$\mu$m at a rate of 500~Hz\,cm$^{2}$.

\subsection{\label{Sec43}Rate dependence of the muon detection efficiency}

\begin{figure}[hbt]
\begin{center}
    \includegraphics[width=0.5\linewidth]{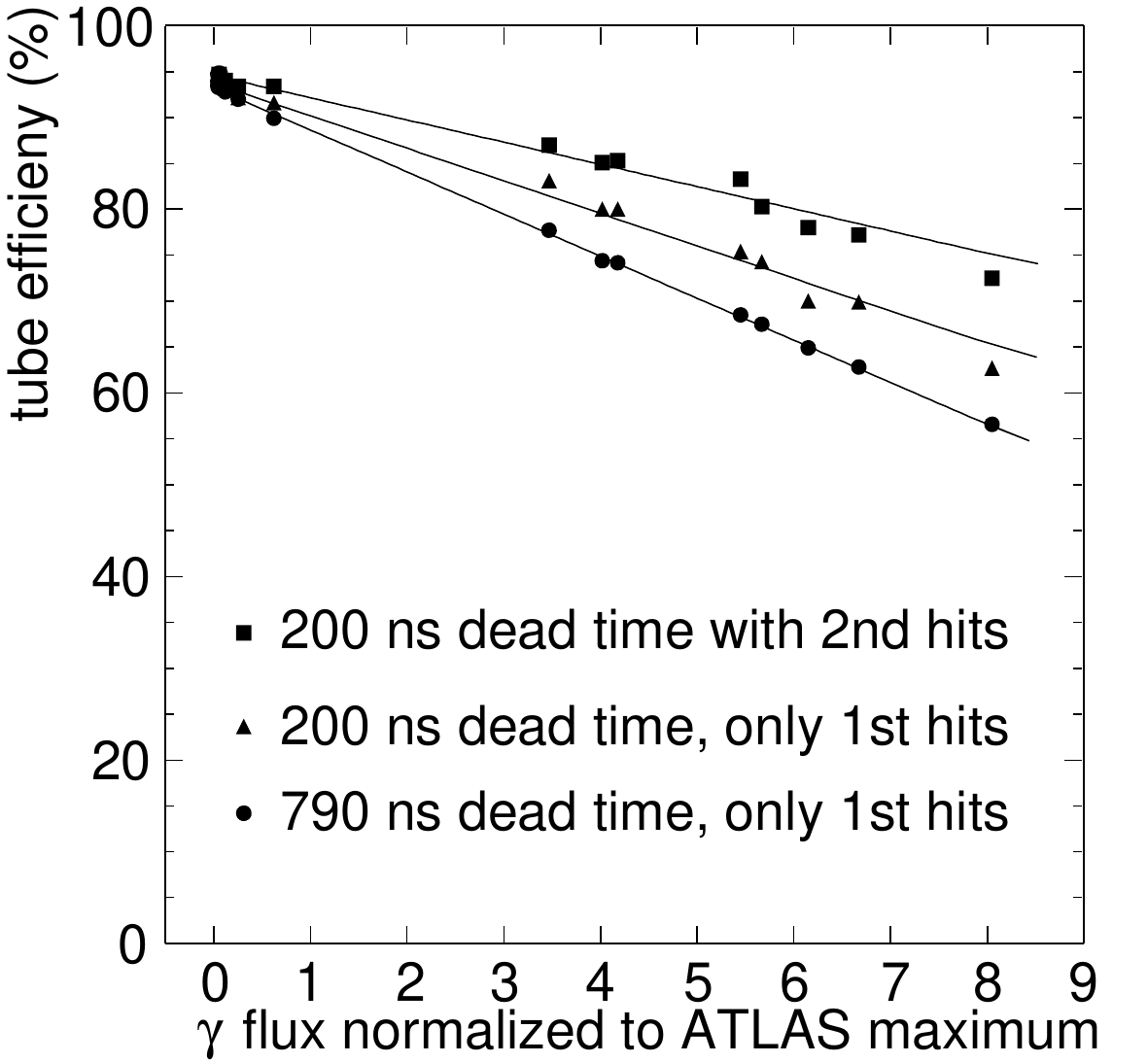}
    \caption{\label{Fig23}$3\sigma$~single-tube efficiency as a function of the background flux for different dead-time settings of the read-out electronics. A flux of 1 corresponds to a background counting rate of 60~kHz per tube and does not contain safety factors acounting for the uncertainty of the rate prediction. The solid lines are straight lines fitted to the data points.}
\end{center}
\end{figure}

High background radiation deteriorates the $3\sigma$~single-tube efficiency because muons traversing a tube within the dead time of the electronics after a background hit are not detected. The $3\sigma$~single-tube efficiency was measured with two settings of the artificial dead time, the nominal value of 790~ns and the minimum of 200~ns. A tube read out with minimum dead time can give more than one hit per trigger because the bipolar signal can cross the discriminator threshold more than once. Muons and photons cause about 1.5~hits per trigger on the average when the minimum dead time is set. The multiplicity is not increased in tubes operated in a magnetic field. The data are analyses in two ways in Figure~\ref{Fig23}:
\begin{enumerate}
    \item   Only the first hit in time is considered. As the minimum dead time of the electronics is much shorter than the maximum drift time of about 700~ns, the single tube efficiency is higher for the minimum dead time than the nominal dead time. The efficiency increase is 2.5\% at the maximum nominal background counting rate of 60~kHz per tube. At a rate of ten times the ATLAS maximum the efficiency gain is 10\%.
    \item   The data with minimum dead time contain seconde hits as mentioned above. Whenever the first hit gives as wrong radius and another hit has been detected in the same tube for the same trigger, the second hit is used. This enables us to recover muon hits which are more than 200~ns after a background hit. As a consequence, the efficiency is further increased. It is greater than 80\% for rates of up to 10~times the maximum rate expected in ATLAS. It should be mentioned, however, that the signal of second muon hits pile up on the preceding background hits. Second hits therefore show a degraded spatial resolution. The resolution is degraded by a factor 1.3-1.4 at the highest rates.
\end{enumerate}
The improvement of the single-tube efficiency due to the reduced dead time and second hits can be exploited by the muon track reconstructino algorithms as shown for occupancy of 35\% in \cite{Primor}. The ATLAS MDT chambers can be efficiently operated at rates of up to about 500~kHz per tube corresponding to an occupancy of 35\%.

\section{Summary}

The performance of ATLAS muon drift-tube chambers has been studied in detail in highly energetic muon beams. Measurements on drift tubes in magnetic fields show that inelastic collisions of drifting electrons on CO$_2$ molecules have to be taken into account in the simulation of the electron drift properties in the Ar:CO$_2$ gas mixture of the MDT chambers. Inelastic collisions are correctly treated by the Garfield simulation programme since version 9 which is hown to provide accurate predictions of the drift properties of the ATLAS muon drift tubes.

Measurements at the Gamma Irradiation Facility at CERN allowed for the measurement of the performance of the MDT chambers in the presence of high $\gamma$~radiation background fluxes. The chambers have a spatial resolution better than 40~$\mu$m at the nominal background rates expected at the LHC design luminosity of $10^{34}$~cm$^{-2}$s$^{-1}$ and a resolution better than 50~$\mu$m for up to five times higher background rates. Efficient muon detection is possible up to background counting rates of 500~kHz per tube corresponding to 35\% occupancy.



\end{document}